\documentclass[aps,pre,preprint,superscriptaddress]{revtex4}
\usepackage{amsmath,amssymb}
\usepackage{graphicx}
\usepackage{dcolumn}
\usepackage{bm}
\newcommand{\eff}{\mathrm{eff}}
\newcommand{\nn}{\mathrm{m}}

\newcommand{\VECr}{{\boldsymbol{r}}}
\newcommand{\lb}{\left}
\newcommand{\rb}{\right}
\newcommand{\tens}[1]{{\mathsf #1}}
\newcommand{\bnabla}{\mbox{\boldmath $\nabla$}}

\begin{document}
\title{Many-body interactions and melting of colloidal crystals}
\author{J. Dobnikar}
\affiliation{Universit\"at Konstanz, Fachbereich Physik, 
D-78457 Konstanz, Germany}
\author{Y. Chen}
\affiliation{Department of Physics, 
Lanzhou  University, Lanzhou  Gansu, 730000, China}
\author{R. Rzehak}
\affiliation{Institut f\"ur Festk\"orperforschung, Forschungszentrum J\"ulich,
D-52425 J\"ulich, Germany}
\author{H.H. von Gr\"unberg}
\affiliation{Universit\"at Konstanz, Fachbereich Physik, 
D-78457 Konstanz, Germany}
\date{\today}

\ \\
\ \\

\begin{abstract}
  We study the melting behavior of charged colloidal crystals, using a
  simulation technique that combines a continuous mean-field
  Poisson-Boltzmann description for the microscopic electrolyte ions
  with a Brownian-dynamics simulation for the mesoscopic colloids.
  This technique ensures that many-body interactions between the
  colloids are fully taken into account, and thus allows us to
  investigate how many-body interactions affect the solid-liquid phase
  behavior of charged colloids.  Using the Lindemann criterion, we
  determine the melting line in a phase-diagram spanned by the
  colloidal charge and the salt concentration. We compare our results
  to predictions based on the established description of colloidal
  suspensions in terms of pairwise additive Yukawa potentials, and
  find good agreement at high-salt, but not at low-salt concentration.
  Analyzing the effective pair-interaction between two colloids in a
  crystalline environment, we demonstrate that the difference in the
  melting behavior observed at low salt is due to many-body
  interactions. If the salt concentration is high, we find
  configuration-independent pair-forces of perfect Yukawa form with
  effective charges and screening constants that are in good agreement
  with well-established theories. At low added salt, however, the
  pair-forces are Yukawa-like only at short distances with effective
  parameters that depend on the analyzed colloidal configuration.  At
  larger distances, in contrast, the pair-forces decay to zero much
  faster than they would following a Yukawa force law.  This is
  explained with a screening effect of the macroions.  Based on these
  findings, we suggest a simple model potential for colloids in
  suspension which has the form of a Yukawa potential, truncated after
  the first coordination shell of a colloid in a crystal.  Using this
  potential in a one-component simulation, we find a melting line that
  shows good agreement with the one derived from the full
  Poisson-Boltzmann-Brownian-dynamics simulation.
\end{abstract}
\maketitle

\newpage


\section{Introduction}
\label{intro}

The classic Derjaguin-Landau-Verwey-Overbeek (DLVO) theory
\cite{verwey,likos01,levin02,reviewBelloni00,rev:attard,rev:loewen} 
predicts that an isolated pair of charged colloidal
spheres in an aqueous salt solution interacts via a repulsive Yukawa
potential at large separations, a prediction that has recently been
confirmed by direct experimental measurements
\cite{crocker352,vondermasson1351,crocker1897}.  However, in nature
colloidal particles usually do not occur simply as pairs, but rather in 
form of suspensions. Then the DLVO pair potential does not always provide 
a valid description of the intercolloidal interactions as becomes
clear from the following considerations. The DLVO potential is an
effective colloid-colloid potential, obtained after integrating out
the microionic degrees of freedom; it is governed by the density
distribution of the ionic fluid between two interacting colloids.
However, when more than two colloids are present, as it is the case in
colloidal suspensions, then this density distribution can be
sensitively influenced by other colloids in the neighborhood, i.e., by
both the presence of the fixed charges on the colloidal surfaces and
the exclusion of electrolytic solution from the volume occupied by the
colloid. Because of these confinement and polarization effects,
colloidal interactions in suspensions in principle are non-additive,
that is, the total potential energy can not be written as a sum of
simple pair-interactions.

The implications of this interesting many-body aspect of
charge-stabilized colloidal suspensions have been addressed by various
authors. Linse \cite{linse91} studied asymmetric two-component
electrolytes (charge asymmetry 1:20) within the primitive model and
explored pair- and triplet-correlations among colloidal particles in
concentrated suspensions. He compared his data with the results of a
simulation of an effective one-component system, consisting only of
colloids interacting via effective pair-potentials, chosen such that
the pair-correlations in the original two-component system and in the
reference one-component system are identical. Although in general
rather similar triplet correlation functions were obtained, considerable 
differences became visible at small distances, pointing to the
existence of many-body forces. In a more direct way, three-body forces
have been analyzed by L\"owen and Allahyarov \cite{allah} using
primitive model simulations in a triangular geometry of colloids.
Interestingly enough, these three-body forces are attractive,
representing a substantial correction to the repulsive pair-forces, a
finding that has been corroborated by recent primitive model \cite{wu}
and Poisson-Boltzmann simulations \cite{russ}, see also \cite{tehver1335}.

The present work addresses the question how three- and higher-order
body forces affect the melting behavior of colloidal crystals.
Starting from crystalline configurations, we compute the mean-square
displacement of colloids about their lattice site and use Lindemann's
rule to estimate the solid-liquid phase-boundaries. The simulation
technique we use, has been suggested by Fushiki \cite{fushiki6700}; it
combines a Poisson-Boltzmann field description for the microions with
a Brownian dynamics simulation for the macroionic colloids, an
approach which ensures that many-body interactions between the
colloids are fully included. In order to identify many-body effects we
have to contrast the phase-diagram resulting from these simulations to
the predictions based on the established description of colloidal
suspensions in terms of pairwise Yukawa potentials. This requires a
procedure by which we can map our original many-body non-linear system
onto an effective linear pairwise additive Yukawa system. This is done
here by calculating effective pair-forces between colloids which are
then fitted to Yukawa pair-potentials. Our main result is that the
melting line of a colloidal system with purely pair-wise additive
potentials is shifted towards the crystalline side of the
phase-diagram if many-body forces are added. In view of the fact that
three-body forces are already known to be attractive, this result is
what one would expect: for a colloidal crystal whose stability results
from repulsive pair-forces, adding attractive triplet forces should
have a destabilizing effect on the crystal so that it melts earlier
than it would without these three-body forces. We will however see
that it is not straightforward to spot the many-body forces as
being responsible for this effect.

After some general remarks in sec.~\ref{sec2} on many body
interactions in colloidal systems, we will give in sec.~\ref{sec3} a
description of the system and the simulation methods used.
Sec.~\ref{sec4} is then devoted to a presentation of our effective
force calculations in colloidal FCC and BCC crystals, followed by
sec.~\ref{sec5} in which we will discuss the phase behavior of a
Yukawa system. This, together with the results from sec.~\ref{sec4},
will enable us to compare in sec.~\ref{sec6} the melting behavior
resulting from our simulations with that of a simple Yukawa system.
Sec.~\ref{sec7} contains a summary of our results and the general
conclusion.

\section{Many-body interactions and density-dependent
  pair-potentials in colloidal systems}
\label{sec2}

There exists a bijective one-to-one mapping between a unique pair
potential $u_{\eff}(r,\rho)$ and the pair correlation function $g(r)$
at the density $\rho$ \cite{henderson74}. That is, even in systems
where the true interaction potential contains a strong three-body
contribution $u_{3}(r_{12},r_{13},r_{23})$, one can interpret the pair
correlation function in terms of (now effective) pair-potentials
$u_{\eff}(r,\rho)$. These effective potentials are to be distinguished
from the true pair-potential $u(r)$. To lowest order in $\rho$,
$u_{\eff}(r,\rho)$ reads \cite{attard92,reatto87,ram77}
\begin{equation}
  \label{eq:1}
  \beta u_{\eff}(r) = \beta u(r)
  - \rho \!\int\! g_{0}(r_{13}) g_{0}(r_{23})
  (e^{-\beta u_{3}(r_{12},r_{13},r_{23})} - 1) \, 
  d\VECr_{3}
\; ,
\end{equation}
where $\beta = 1/k_{B}T$ and where $g_{0}(r)$ is that radial
distribution function which is based on the true pair potential $u(r)$
alone. Eq.~(\ref{eq:1}) states in essence that the three-body
interactions can be integrated out, and that the resulting effective
pair-potential becomes state-dependent. In other words, taking
eq.~(\ref{eq:1}) as a model pair-potential in a system with pair-wise
interactions only, one can, at low density, calculate the correct
pair-correlation function for the true system, consisting of particles
interacting via $u(r)$ and $u_{3}(r_{12},r_{13},r_{23})$.

Eq.~(\ref{eq:1}) is the key to an understanding of a recent experiment
in 2D suspensions of charged colloids \cite{klein,brunner} which being
intimately related to the subject of this paper, we briefly summarize
in the following. In this experiment digital video microscopy has been
used to systematically investigate the effective pair-potentials
$u_{\eff}(r,\rho)$ as a function of the colloid densities $\rho$. At
relatively low densities, potentials very close to Yukawa form have
been observed, just as standard DLVO theory predicts.  For increasing
density, however, the measured potentials were Yukawa-like only at
very short distances but showed clear and systematic deviations from
the Yukawa form at larger distances.  From eq.~(\ref{eq:1}) it is seen
that a density-dependence of $u_{\eff}(r,\rho)$ reveals the existence
of many-body interactions, in particular three-body interactions.
According to \cite{allah,wu,russ} three-body forces between charged
colloids are attractive, so $u_{\eff}(r)$ in eq.~(\ref{eq:1}) should
be smaller than $u(r)$ in the distance regime where
$u_{3}(r_{12},r_{13},r_{23})$ is appreciable.  This has indeed been
observed in \cite{klein,brunner}.  Another observation made in this
experiment has been that the radial distance where the measured
$u_{\eff}(r,\rho)$ started to show deviations from $u(r)$ was correlated
with the mean colloid-colloid distance $d_{m}=\rho^{-1/2}$ (in 2D).

These observations show that three-body interactions between the
colloids in the suspension are non-negligible and that they indeed
seem to be attractive as predicted in \cite{allah,wu,russ}. A clearer
idea of the experimental findings is given by the following
interpretation: two colloids, A and B, separated by a distance
$r<d_{m}$, will behave as if they were in isolation, i.e. the
interaction is determined solely by the microions and the
pair-potential is Yukawa-like.  However, if $r \ge d_{m}$, then it is
very likely that there is a third colloidal particle, C, near the line
joining the two interacting colloids A and B.  This third particle has
a threefold effect on the interaction between the particles A and B.
First, there is a confinement effect; the electrolyte solution is
excluded from the volume occupied by particle C.  Secondly, colloids
are usually made of materials that have a much lower dielectric
constant than that of the aqueous solvent; electric field lines will
then also be excluded from the volume of particle C.  And, finally,
the colloids are charged; thus the charges of particle C will affect
the microionic charge distribution between the two interacting
colloids A and B. All three effects result in a pair-interaction
between A and B that will depend on the position of particle C. This
can be taken into account by choosing a description either in terms of
density-dependent effective pair-potentials or in terms of
density-independent pair-potentials plus many-body interactions.  The
effect of particle C on the interaction between A and B is the
stronger the nearer C is to the joining line between A and B, and
certainly strongest, if C lies directly on this line. In \cite{russ},
this coaxial geometry has been considered, and it has been found that
the three-body potential equals a considerable fraction (up to 90\%
for almost touching colloidal spheres) of the negative of the
pair-potential between the two outer particles (particle A and B).
This means that the middle colloid essentially shields (or
''screens'') the two outer ones from each other; it blocks the direct
(repulsive) pair interaction between them.  In other words, the
triplet interaction describes, to lowest order, the screening of pair
interactions due to additional macroions. One may say that the
interaction between two colloids at large distances ($r \ge d_{m}$) is
screened by both microions and macroions, while it is screened just by
the microions if $r<d_{m}$; so the effective $\kappa$ in the former
case is larger than in the latter case because the ionic strength is
higher when macro- and microions are counted in comparison to the case
where only the microions are taken into account \cite{warren}. Because
the screening mechanism involving macro- and microions is much more
efficient, the pair interaction between colloids in suspension will
decay much faster than the interaction between an isolated pair of
colloids.

In fact, it is just this cross-over from a microion-dominated to a
macroion-dominated screening that has been observed in the experiment
in \cite{klein,brunner}, and the distance where this happens was
indeed roughly given by $d_{m}$, beyond which value the effective 
pair-potentials decayed so rapidly to zero, as if they were cut off. 
Such a ''cut-off'' like behavior, caused by macroion screening, is also 
found in the effective force curves we present in Sec.~\ref{sec4}.  It
also plays a decisive role when in Sec.~\ref{sec5} we try to
understand the solid-liquid phase behavior of colloidal suspensions
with truncated and density-dependent effective pair-potentials, chosen
to model the effect of many-body interactions.

In rare-gas systems it has long been known that a description in terms
of pair-potentials is not strictly valid because the interaction
between two particles is again disturbed by the presence of a close
third particle \cite{henderson76}. In these systems, one usually
describes the interaction between the atoms by combining a
pair-potential with the so-called Axilrod-Teller triple-dipole
potential \cite{axilrod43}. Mainly the effect of this additional
many-body potential on the pair-correlations functions has been
studied for krypton \cite{tau89,jakse02}, argon
\cite{barker71,bukowski01} and xenon \cite{bomont02}. Most of these
papers are large-scale computer simulations, including the classic
work of Barker et al.  \cite{barker71}, others present perturbation
\cite{ram77} and integral-equation theories \cite{attard92,bomont02}
in which the pair-correlation functions is calculated from the
Ornstein-Zernike equation using a closure that involves the effective
pair-potential from eq.~(\ref{eq:1}). Concerning the gas-liquid
equilibrium, the effect of the Axilrod-Teller interaction potential
has been studied for argon in \cite{bukowski01}, where only a moderate
difference to the pairwise interactive case was found.

Contrary to these rare-gas fluids, where one is simply stuck with the
interaction dictated by the electronic structure of the atoms, the
interactions between charged colloidal particles can be adjusted by
adding salt to the electrolyte. Specifically, three-body interactions
between the colloids can be 'switched-off' just by adding salt.  This
allows to investigate experimentally how the behavior of a suspension
of charged colloids depends on their interactions.  The possibility of
tuning the interactions in an experiment makes colloidal suspensions a
system that is ideally suited for studying many-body effects.

The salt concentration thus is an important quantity in the following
discussion. It is best specified by means of the inverse Debye
screening length $\kappa^{-1}$. This is also a natural measure for the
interaction range of effective colloidal interactions, as it gives the
thickness of the spherical double-layer around a single isolated
colloidal sphere.  Comparing $\kappa^{-1}$ to the mean colloid-colloid
distance $d_{m}$, we can estimate when we expect three-body
interactions to be important. If two spherical double-layers overlap,
the screening of the colloidal charges becomes incomplete and the
charge distributions on the colloids involved begin to interact. If
the colloid density is low, i.e. if $d_{m}\kappa \gg 1$, it is obvious
that such an overlap will almost always occur for pairs of colloids
only. At high colloid densities, however, i.e. if $d_{m} \kappa \sim
1$, there is a high probability that more than one other colloid is
within the range of the double-layer $\kappa^{-1}$ around any
colloidal particle.  Many-body forces between the colloids thus become
important under low-salt conditions and/or at high colloid densities
\cite{warren}. 

Unlike in rare-gas systems, the effect of triplet interactions
\cite{allah,wu,russ,tehver1335} on the phase-behavior of charged
colloids has not received much attention. Wu et al. \cite{wu} have
used their simulated three-body potentials to calculate the densities
of the coexisting phases by simple van-der-Waals theories for the
solutions and the colloidal crystals. They studied colloidal systems
in monovalent electrolyte solutions, with and without triplet forces,
and found only small differences. However, they investigated systems
at rather high salt concentrations ($0.05$ M) while as pointed out
above, the physics of charged colloidal suspensions becomes special
only at low-added salt \cite{beresfordsmith}.  The interesting
features of phase equilibria in charged colloidal suspensions at
low-salt conditions have also been pointed out in \cite{vanroij99}.
Linse et al., in a series of papers
\cite{linse91,linse4359,linse4208,linse3917}, have explored by
primitive model simulations the structure, phase-behavior and
thermodynamics of a model system consisting of charged colloids and
point counterions (with up to 80 counterions per colloid); these
authors concentrated mainly on a possible gas-liquid
phase-coexistence, but not on the crystallization behavior of their
system.

The considerations made above show that in the presence of many-body
interactions, the effective pair-potential becomes density-dependent.
That a density-dependence of the pair-potential can produce a
van-der-Waals like instability even in fluids with purely repulsive
pair potentials has been shown for a colloidal system by Dijkstra et
al.\cite{dijkstra1219}. It can also lead to a second liquid-liquid
phase separation in systems that already show a gas-liquid
phase-coexistence \cite{almarza2038}.  A broad discussion of
density-dependent pair-potentials with many illustrative examples of
typical soft-matter systems is given by Louis \cite{louis02},
including recent theoretical studies of coarse graining of polymers as
'soft colloids' \cite{louis00,bolhuis01}.  Louis points out that the
effective density-dependent pair potential in eq.~(\ref{eq:1}), though
reproducing the correct pair-structure, will not generate the correct
virial pressure. The implications of this inconsistency for the
thermodynamics of the fluid is considered in \cite{tejero02}. This
shows the importance of investigations like ours taking full account
of the many-body nature of colloidal interactions.

\section{Simulation technique}
\label{sec3}

We now give a short description of the technique we have used to
calculate effective forces in sec.~\ref{sec4} and the melting line of
colloidal crystals in sec.~\ref{sec6}.  More details about our
simulation method can be found in \cite{technical_paper}.

We consider a fluid of highly charged identical colloidal spheres
suspended in a structureless medium of dielectric constant
$\varepsilon$ at temperature $T$.  The Bjerrum length characterizing
the medium is defined as $\lambda_{B} = e^{2}\beta/\varepsilon$, with
$e$ the elementary charge.  The colloidal spheres have radius $a$,
charge $-Ze$, and thus a surface charge density $-e\sigma = -Ze/4 \pi
a^{2}$.  The positions of their centers are denoted by ${\bf R}_i$
($i=1\ldots N)$.  The suspension is assumed to be in osmotic
equilibrium with an electroneutral and infinite reservoir of
monovalent point-like salt ions with total particle density $2c_{s}$
and inverse Debye screening length $\kappa=(8\pi\lambda_B
c_{s})^{1/2}$, and also in thermal equilibrium with a heat bath at
constant temperature. In the Poisson-Boltzmann (PB) mean-field
approach, the density distribution of the positive and negative
microions in the region $G$ exterior to the colloidal spheres at
$\{{\bf R}_i\}$ is given by $n_{\pm} = c_{s} e^{\mp \phi(\VECr)}$
where $\phi(\VECr)$ is the normalized electrostatic potential
satisfying the PB equation
\begin{equation}
  \label{eq:2}
\begin{array}{rclcl}
 \bnabla^{2} \phi(\VECr) & = & \kappa^{2} \sinh \phi(\VECr)\:, 
& \quad  &\VECr \in G \\
{\bf n}_{i}\bnabla \phi & = & 4 \pi \lambda_{B} \sigma& &\VECr
\in \partial G_i, \: i=1,\ldots,N \:,
\end{array}
\end{equation}
with $\partial G_i$ being the surface of the $i$-th colloid with
outward pointing surface normal ${\bf n}_{i}$.  Constant-charge
boundary conditions are assumed for all $N$ colloid surfaces.

Eq.~(\ref{eq:2}) is based on a description in the semi-grand-canonical
ensemble. The suspension is assumed to be separated from the reservoir
by a semi-permeable membrane, through which solvent molecules and
microions can pass, but not colloidal particles. This leads to an
imbalance in the osmotic pressure across the membrane.  The
equilibrium between the suspension and the salt reservoir is referred
to as Donnan equilibrium \cite{Don24,Ove56,hill56}.  The effective
average salt concentration $n_{s}$ in the suspension is not the same
as the reservoir salt concentration $c_{s}$, but is a function of the
physical parameters of the system and can be calculated, e.g., in the
PB cell model \cite{reus97,TaLe98,deserno}. Here, we can compute
$n_{s}$ from the total number of microions, obtained by integrating
the microionic density distribution, $n_{+}(\VECr) + n_{-}(\VECr)$,
over $G$. Since the actual salt concentration in the system, $n_{s}$,
is of no interest in the following considerations we dispense with
quantifying it, but will instead specify just $c_{s}$, or,
equivalently, $\kappa a = (8 \pi (\lambda_B/a) c_{s}a^{3})^{1/2}$
which for simplicity we will refer to as 'salt concentration'.

In many experiments, the system is actually not coupled to a salt
reservoir and only $n_{s}$ is known. One can then imagine it to be
coupled to a (physically not existing) reservoir with a salt
concentration $c_{s}$ chosen such that it corresponds to the given
$n_{s}$ in the system. Practically, this amounts to inverting the
function $n_{s}(c_{s})$. Provided we only choose the right $c_{s}$
corresponding to the known $n_{s}$, a real reservoir when coupled to
the system would leave the microion density distribution in the system
completely unaltered. In other words, every system, be it isolated or
not, can be treated as if it {\em were} in contact with a reservoir.
This implies that all our considerations presented here in the
semi-grand canonical ensemble are also valid for experiments in the
canonical ensemble in which the system is actually not coupled to a
salt reservoir.

The PB equation \cite{rev:andel} can be derived from a mean-field free
energy functional
\cite{Lev39,BrRo73,ReRa90,trizac97,rev:deserno,tama02} 
which itself is obtained from the underlying Hamiltonian either 
(i) as the saddle point of the field theoretic action 
\cite{PoZe88,CoDu92,netz00,BoAn00}, 
(ii) as a density functional reformulation of the partition function 
combined with a first order cumulant expansion of the correlation term
\cite{lhm}, or (iii) from the Gibbs-Bogoljubov inequality applied to a
trial product state \cite{rev:deserno}. The PB approximation gives a
good description in the limit of high temperature and/or small charge
densities (weak-coupling limit) when the microionic correlations that
are neglected at the mean-field level are unimportant. Indeed,
these microion correlations play only a negligible role in our system
containing only monovalent ions, as has been shown in various studies
\cite{groot,valleau,lhm,lobaskin99,lobaskin01}.  Groot \cite{groot},
for example, quantified the validity of the PB approach by comparing
it to a cell model Monte-Carlo simulation and showed that the
deviations are already tiny for a ratio $\lambda_{B}/a = 0.03$.  Since
we are here considering an even smaller value $\lambda_{B}/a=0.012$,
the PB approach is perfectly justified.

To simulate now the dynamics of the large macroions suspended in the
microionic liquid, the PB field description of the small ions in
eq.~(\ref{eq:2}) is combined with a Brownian dynamics simulation of
the macroions, similar to Refs.~\cite{fushiki6700,gilson95}.  For each
spatial configuration of the macroions, the electrostatic potential
$\phi$ arising from all charges in the system is determined by solving
eq.~(\ref{eq:2}) with boundary conditions determined by the positions
of the macroions and the charges on them.  To truncate the ideally
infinite suspension volume with a minimum of finite size effects, a
cubic simulation box with periodic boundary conditions on its faces is
considered.  From the electrostatic potential, the force
${\bf F}^{\phi}_i$ acting on each of the macroions is calculated by
integrating the normal component of the stress tensor
\begin{equation}
\label{eq:3}  
\beta\tens{T} =
\frac{1}{8 \pi \lambda_B}
\bigg(  \Big(   2 \kappa^{2} \lb( \cosh \phi - 1 \rb)
              + \lb( \bnabla \phi \rb)^{2} 
       \Big)\tens{1}
     - 2 \bnabla \phi \otimes \bnabla \phi 
\bigg)
\end{equation}
over a surface enclosing the respective macroion.  These forces are
then used to perform one time step in a Brownian dynamics simulation
for the motion of the macroions, generating another colloidal
configuration for the next iteration cycle.

For large colloidal charge $Z$, the solution of the PB equation may
develop rather steep boundary layers close to the particle surfaces
while far away it varies only gradually.  To provide an accurate
resolution of the boundary layers with a reasonable number of grid
points, we follow Ref.~\cite{fushiki6700} and define a spherical grid
in a shell centered about each of the macroions overset on a Cartesian
grid covering the simulation box.  The resolution near the particle
surfaces is further enhanced by choosing the radial spacing of
spherical grid points such that the inverse radial coordinate has a
constant increment \cite{muller28}.  The solution of the PB equation
on the domain $G$ outside the macroions is then obtained in four
steps: First the PB equation is solved in each of the $N$ spherical
shells $\Omega_{1}, \ldots, \Omega_{N}$ assuming given potential
values at their outer edges.  Interpolation of these spherical shell
solutions to the Cartesian grid next yields boundary values for the
interstitial region $\Omega_{0}$ which is the part of the domain $G$
not contained in any of the spherical shells.  With these boundary
values the PB equation is then solved in the interstitial region.
Finally, new boundary values on the outer edges of the spherical
shells are found by interpolating back from the interstitial solution.
These steps are repeated until a converged solution on the whole
domain $G$ is obtained.

The PB equation is discretized by a finite volume procedure
\cite{fletcher91} to handle the singular points on the polar axis of
the spherical grids.  For the non-singular grid points of the
spherical and Cartesian grids this reduces to a simple second order
finite difference method. On each of the grids the discretized PB
equation is solved by a nonlinear 'Successive Over-Relaxation' (SOR)
method \cite{ortega70}.  The iteration is terminated when the maximum
of the absolute value of the residuum for all grid points drops below
some desired accuracy.  Note that it is not necessary to iterate to
convergence in each of the repeatedly solved sub-domain problems for
$\Omega_{0}, \ldots, \Omega_{N}$.  Only in the last pass of this
repetition, the full convergence must be ensured.

The integration of the normal component of the stress tensor that
gives the force on the macroions turns out to be best performed in the
middle of the spherical shell $\Omega_{p}$ around each particle.  The
derivatives of the electrostatic potential appearing in the stress
tensor are discretized by central differences while the midpoint rule
is used to approximate the surface integral.

The equation of motion for the macroions reads
\begin{equation}
\label{eq:4}
\zeta\dot{\bf R}_i
={\bf F}^{\phi}_i
+ \sqrt{2 k_{\rm B} T \: \zeta \,} \: \mbox{\boldmath $\xi$}_i
\;,
\end{equation}
where $\zeta = 6 \pi \eta a$ is the Stokes friction coefficient of a
macroion in the solvent of viscosity $\eta$, and where $\vec{{\xi}}_i$
is an uncorrelated Gaussian white noise with zero mean and unit
variance. This equation is advanced in time by a stochastic Euler
algorithm \cite{kloeden92,oettinger96} with time step $h$.  A
time-discrete approximation to the white noise $\vec{\xi}_i$ is
furnished by a sequence of random numbers with zero mean and variance
$h$ drawn independently at each time step.  Since we are interested
only in calculating averages from the trajectories, only the first two
moments of the distribution of these random numbers are important
\cite{kloeden92,oettinger96}.  

This method allows to calculate structural and thermodynamic
properties of charge-stabilized colloidal suspensions in the parameter
regime of large macroionic charge, low salt concentration, and
relatively high colloid volume fraction, where many-body interactions
between the macroions become important. By describing the small ions
in terms of a continuous density, the computational effort becomes
independent of their number.  In this way, the main limitation, which
precludes the application of primitive model simulations to the above
parameter regime is circumvented.  At the same time, in contrast to
analytical treatments like linearized DLVO theory or cell models,
macroionic many-body effects are fully accounted for in our approach.

There are a few studies \cite{loewen673,lhm,tehver1335,fushiki6700}
where similar simulations have been carried out. The method used by
L\"owen et al. \cite{lhm,loewen673} is based on a density functional
approach which includes also a LDA micro-ion correlation term; it
becomes equivalent to the method we used, if one neglects the
microion-microion correlations.  The technically most difficult part
of these simulations are the numerical solution of the boundary-value
problem in eq.~(\ref{eq:2}) for arbitrary $\{{\bf R}_i\}$ which is
solved in different ways in \cite{lhm,loewen673,tehver1335} and
\cite{fushiki6700}.  The overset-grid technique we here use following
\cite{fushiki6700}, is more accurate for a comparable number of grid
points than the single grid computations used in
\cite{lhm,loewen673,tehver1335}, and thus allows to consider more
highly charged colloids. L\"owen et al. \cite{lhm} and Fushiki
\cite{fushiki6700}, both working in the salt-free case, used this
method to calculate structure functions and pair distribution
functions of colloidal dispersions in the fluid phase, compared with
MC simulations, and predictions of the DLVO theory and the PB cell
model \cite{alexander}. Mapping many-body interactions onto optimal
Yukawa pair-interactions is the key idea of \cite{loewen673},
considering, as in the present paper, colloidal systems with added
salt.

We have four different length scales in the system which are: the
screening length $\kappa^{-1}$, the colloidal hard-sphere radius $a$,
the Bjerrum length $\lambda_{B}$, characterizing the solvent, and the
mean-distance $d_{m}=\rho^{-1/3}$ (in 3D) between the colloids in a
suspension at a number density $\rho$ (volume fraction $\eta= 4 \pi
\rho a^{3}/3$).  A further physical parameter of our system is the
colloidal charge $Z$.  Taking (i) the radius $a$ of the colloidal
spheres as unit-length scale and concentrating (ii) on colloid
particles of a typical size of $a = 60$nm common in experimental
studies \cite{gast89,yama5806} in aqueous systems at room temperature
where $\lambda_{B}=7.2\:\AA$, we fix the ratio $\lambda_{B}/a =
0.012$, a value which is used throughout this work.  The state space
of our colloidal system is thus three-dimensional, spanned by $\kappa
a$ (salt concentration), $Z$ (colloidal charge) and $d_{\nn}/a$
(colloid density), or, equivalently, the volume-fraction $\eta=4 \pi
(a/d_{\nn})^{3}/3$.  A point $\vec{X}$ in state space is then defined
by $\vec{X} \equiv (\kappa a, Z, \eta)$.  Since a systematic
exploration of the full 3D state space is computationally too
expensive, we here focus on a 2D cut at large volume fractions and
take, if not otherwise specified, $\eta = 0.03$ , a high but
experimentally realistic value \cite{yama5806} for which we expect
pronounced many-body effects. The number of colloids placed in the
simulation box is $N=54$ or $N=108$ for calculations starting from BCC
or FCC configurations, respectively. No substantial differences have
been observed using only 32 particles in the FCC case, so finite-size
effects should be negligible.

\section{Effective force calculations}
\label{sec4}

Effective colloid-colloid pair forces are needed in sec.~\ref{sec6} to
understand and appreciate the meaning of our simulated phase-diagram.
In this section we present results for such effective pair-forces,
obtained for colloids in a crystalline environment which has either
FCC or BCC symmetry. Choosing two particles A and B in such a crystal,
we first calculate the total force ${\bf F}_{B}^{1}$ acting on
particle B with particle A present.  This is done by solving the PB
problem for the given configuration and by then calculating ${\bf
  F}_{B}^{1}$ with the help of the stress-tensor integration, as
described in sec.~\ref{sec3}.  Then particle A is removed leaving all
the other particles in place and once again the total force
${\bf F}_{B}^{0}$ acting on particle B is calculated.
The force exerted by particle A on particle B is just the difference
between the two.  This procedure is then repeated many times for
distorted crystalline configurations in which all particles are on
their lattice sites except particle A which is displaced by some
distance. This finally results in the effective force curve:
\begin{equation}
\label{eq:5}
{\bf F}_{AB}(\VECr) =
{\bf F}_{B}^{1}(\VECr)-{\bf F}_{B}^{0}
\end{equation}
where $\VECr = {\bf R}_{A} - {\bf R}_{B}$.  Projecting ${\bf F}_{AB}$
onto the axis connecting particle A and B, we obtain the force
component
\begin{equation}
\label{eq:6}
f(\VECr) = \beta a {\bf F}_{AB}(\VECr) \frac{\VECr}{|\VECr|}
\end{equation}
made dimensionless here by multiplication with $\beta$ and $a$.  It is
clear that if the true interactions in the system are pairwise
additive, the effective interaction, resulting from this procedure, is
by construction identical to the true pairwise interaction potential.
This means, in particular, that the effective interaction is
independent of the direction of $\VECr$ and also independent of the
arrangement of the surrounding particles (all particles but A and B).
On the other hand, if the true interactions between the particles are
many-body in nature, then this procedure folds all these many-body
interactions into the resulting pair interaction, similar to what is
done in eq.~(\ref{eq:1}).  This pair interaction will then depend on
the crystalline environment and on the direction in which particle A
is displaced.

An effective force curve under high-salt conditions, $\kappa a = 2.0$,
is shown in Fig.~(\ref{fig1}). Then, $\kappa^{-1}$ is $ 0.5 a$ and is
thus much smaller than the mean-distance at $\eta=0.03$ which is
$d_m=5.2a$; so pairwise additivity is expected. Indeed, we obtain the
same force curve, regardless which direction we choose for the
displacement of A and which of the two crystalline configurations we
examine. In both configurations we move particle A along highly
unsymmetrical directions in order to avoid bumping into other
particles.  The pair-force we expect to find in this high-salt situation
is a DLVO Yukawa pair-force
\begin{eqnarray}
\label{eq:7}
f_{\mathrm{DLVO}}(r) & = & b(r) e^{-\kappa_{\eff}r} \\
\label{eq:8}
b(r) & = & 
\left[\frac{Z_{\eff} e^{\kappa_{\eff} a}}{1+\kappa_{\eff} a} \right]^{2}
\frac{\lambda_{B}}{a} \frac{1+\kappa_{\eff} r}{(r/a)^{2}}
\end{eqnarray}
with the effective (or renormalized) screening length and charge,
$\kappa_{\eff}$ and $Z_{\eff}$, commonly introduced into the DLVO
theory to capture effects arising from the non-linearity of the PB
equation \cite{reviewBelloni97}. Both quantities $\kappa_{\eff}$ and
$Z_{\eff}$ are functions of $\vec{X}$ and approach $\kappa$ and $Z$
only in certain limiting cases. Multiplying force curves by
$(r/a)^{2}/(1+\kappa_{\eff} r)$ and plotting them logarithmically, a
Yukawa pair-force appears as a straight line with slope
$-\kappa_{\eff}$. This is a convenient way to check whether a given
pair-interaction is Yukawa-like.  The straight line we observe in the
inset of Fig.~(\ref{fig1}) proves that this is indeed the case
for the interaction we obtain at high-salt concentration. From a fit, 
the effective force parameters are obtained as $Z_{\eff}=1080$ and 
$\kappa_{\eff}=1.98$.
The solid line in the main figure of Fig.~(\ref{fig1}) shows
$f_{\mathrm{DLVO}}(r)$ from eq.~(\ref{eq:7}) using these two values
for $Z_{\eff}$ and $\kappa_{\eff}$.

At low salt concentration, $\kappa a = 0.2$, so that $\kappa^{-1}=5a
\sim d_{m} = 5.2 a$, the interaction we obtain shows deviations from a
Yukawa form as becomes evident from Fig.~(\ref{fig2}).  Only at small
distances a Yukawa-like behavior is observed, but around the mean
distance $d_{m}$ systematic deviations start and develop into a
cut-off at a distance $r \approx 1.9 d_{m}$.  Following the discussion
in sec.~\ref{sec2}, this cut-off is caused by the screening effect of
the other macroions.  Fig.~(\ref{fig3}) summarizes the salt dependence
we observe in going from Fig.~(\ref{fig1}) to (\ref{fig2}).  The force
curves $f(r)$, calculated for four different salt concentrations
($\kappa a = 0.1,0.2,1.0,2.0$), are divided by $b(r)$ from
eq.(\ref{eq:8}), the logarithm is taken and the result divided by
$k_{\eff}$. Then a Yukawa pair-force, following eq.~(\ref{eq:7})
appears as a straight line with the slope $-1$ (solid line in
Fig.~(\ref{fig3})). The values of the parameters $\kappa_{\eff}$ and
$Z_{\eff}$ are obtained by fitting the effective force curves to a
Yukawa potential at small distances where the force is still
Yukawa-like. Fig.~(\ref{fig3}) reveals that the cut-off feature,
observable under low salt conditions ($\kappa a=0.1, \, 0.2$),
vanishes at high ionic strength ($\kappa a=2.0$) where perfect
agreement with the predictions of DLVO theory is observed at all
distances probed. This observation is consistent with the calculations
made in \cite{russ} where attractive three-body forces have been shown
to become appreciable in the salt regime $\kappa a < 1$. Note in
Fig.~(\ref{fig3}) that the pair-interaction at high-salt seems to be
longer ranged than that at low-salt, which of course is just due to
our scaling; the dominant factor determining the range of interaction
is still the salt concentration.

Another effect of the many-body interactions comes to light if one
compares effective force curves based on different colloidal
configurations. Fig.~(\ref{fig4}) compares the force curves for three
different combinations of $Z$ and $\kappa a$, each calculated in both
a FCC and a BCC crystal.  No configuration dependence is observed for
the high salt case (lower and middle pair of curves), while at low
salt (upper pair) the forces clearly depend on the configuration,
i.e., on the crystalline environment.

It has been repeatedly stressed in literature
\cite{linse91,linse91b,beresfordsmith,fushiki88a,belloni519,loewen673}
that the functional form of eq.~(\ref{eq:7}) may be useful even in
those regions of the state space where the pair-potential of DLVO
theory deteriorates. Then the prefactor and the screening length are
treated as fitting parameters to reproduce a given $g(r)$ so that the
connection between the force parameters and the system variables is
lost. For this reason, it is interesting to check how our values for
$Z_{\eff}$ and $\kappa_{\eff}$ compare with the known theories on
effective charges \cite{reviewBelloni97,levin02,Quesada}.  The
pioneering work on colloidal charge renormalization is the paper by
Alexander et al.  \cite{alexander} whose method is based on the
Poisson-Boltzmann cell model. The most efficient way to calculate
Alexander's effective charge has been described by Bocquet et al.
\cite{trizac02,trizac02a}: Solving the PB equation within the
spherical cell model, one obtains the potential $\phi_{R}$ at the cell
edge located at $r=R$ and thus the effective screening factor
$\kappa_{\eff}^{2} = \kappa^{2} \cosh\phi_{R}$ which is used to
calculate Alexander's effective charge \cite{trizac02},
\begin{equation}
  \label{alex}
Z_{\eff} = \frac{\gamma_{0}}{\kappa_{\eff} \lambda_{B}}
\Big[ (\kappa_{\eff}^{2} a R - 1)\sinh(\kappa_{\eff}(R-a))
+ \kappa_{\eff}(R-a) \cosh \kappa_{\eff}
(R-a) \Big]  \:.
\end{equation}
with $\gamma_{0} = \tanh \phi_{R}$. The effective charges and inverse
screening lengths obtained from Alexander's prescription are compared
in Fig.~(\ref{fig5}) with the $Z_{\eff}$ and $\kappa_{\eff}$ that we
obtain from fitting our force curves to eq.~(\ref{eq:7}) at $r\le
d_{\nn}$, that is, in the range where the force is Yukawa-like.  The
predictions of the cell model agree quite well with our results in the
high-salt case ($\kappa a = 2.0$), but not under low salt conditions
($\kappa a = 0.2$). While $Z_{\eff}$ derived from the BCC
configuration is still in good agreement with the results of the cell
model calculation, those effective charges derived from the FCC
configuration are far off the values predicted by the cell model.
Regarding $\kappa_{\eff}$, both the FCC and the BCC based results
disagree with the cell model predictions at $\kappa a =0.2$.  Perhaps,
the most interesting outcome of these calculations is that we find a
pair-force which depends on the colloidal configuration, an
observation which has also been made in a similar system by L\"owen
and Kramposthuber \cite{loewen673}. The relative difference between
the FCC and BCC results for the effective quantities increases
systematically upon reducing $\kappa a$ and/or increasing the bare
charge $Z$ as is evident from Fig.~(\ref{fig6}).

To summarize the main results of this section: if the salt
concentration is high, i.e., if $\kappa^{-1}<d_{m}$, we find
configuration-independent pair-forces of perfect Yukawa form with
effective parameters that are in nice agreement with well-established
theories. On the other hand, in the low-salt case where $\kappa^{-1}
\sim d_{m}$, the effective pair-forces (i) are Yukawa-like only at
short distances with effective parameters that depend on the analyzed
colloidal configuration, and (ii) decay much faster to zero at larger
distances ($r \ge d_{m}$) than they would following a Yukawa
pair-force. Both features, the ''cut-off'' like behavior as well as
the configuration dependence, are clearly a fingerprint of
non-negligible many-body interactions which using our procedure we
have folded into the pair interaction.

The ''cut-off'' feature is particularly interesting as it matches the
observations made in the experiment in \cite{klein,brunner}.  A more
physical explanation of this effect follows from our discussions in
sec.~\ref{sec2}: If there is enough salt in the system, then the
interaction between two colloidal particles has already decayed to
zero at $r\ge d_{m}$, i.e., at distances where other macroions could
start influencing the interaction. Then, the interaction is just a
proper pair-interaction and has a Yukawa form with a screening
behavior that is governed exclusively by the microions in the system.
However, if there are not enough microions in the system to screen the
pair-interaction down to zero at $r \sim d_{m}$, then the influence of
the macroions on the screening has to be taken into account. The
interaction then is a Yukawa-like pair-interaction only at short
distances, but has a many-body character at larger distances which
when folded into a pair-force results in what we call here
''cut-off'' behavior.

We finally should remark that our effective force calculations could
be improved by extracting the force between two colloids with all the
surrounding particles not kept fixed at their positions in a
crystalline configuration as in this calculation but instead
undergoing thermal motion.  Especially close to melting, the effective
force thus obtained would incorporate the many-body interactions
averaged over many different colloidal configurations; a possible
configuration- or direction- dependence then, of course, would be no
longer visible.  In the light of the results in
\cite{linse91,linse91b,belloni519,loewen673} it is not unlikely that
the resulting force can again successfully be fitted to a Yukawa
potential.

\section{The solid-liquid phase behavior of a simple Yukawa fluid}
\label{sec5}

The freezing and melting of charge-stabilized colloidal suspensions
has been studied extensively by representing the colloidal system as a
simple Yukawa fluid; many excellent review articles are available
\cite{loewen94,arora98,chak,kuhn97,wang99}.  The phase-behavior has
been investigated analytically \cite{chak,tejero96,lukatsky00,wang00}
and by computer simulation
\cite{rosenberg87,rkg,frenkel,dupont93,meijer97,azhar00}.  The solid
phase is BCC ordered in the region of low $\kappa$ and FCC ordered at
high $\kappa$. At higher temperature there is a disordered fluid
phase. We here concentrate just on the melting line. In all these
studies, the particles interact via the DLVO pair potential
\begin{equation}
\label{eq:10}
\beta u_{\mathrm{DLVO}}(r)  = 
\left[\frac{Z_{\eff} e^{\kappa_{\eff} a}}{1+\kappa_{\eff} a} \right]^{2}
\lambda_{B} \frac{e^{-\kappa_{\eff}r}}{r} \:.
\end{equation}
In this section we accept this picture, treat the suspension as a
simple liquid and make some remarks about its phase-behavior which are
necessary for the considerations presented in the next section.

The phase-diagram of a Yukawa system has first been calculated by
Robbins, Kremer and Grest (RKG) \cite{rkg} using the Lindemann melting
criterion \cite{lindemann} which states that a crystal begins to melt
when the rms displacement in the solid phase is a fraction of 19 \% of
$d_{\nn}$.  For a system of point-like Yukawa particles, interacting
via
\begin{equation}
\label{eq:11}
u(r) = U_{0} \frac{e^{-\lambda r/d_{m}}}{r/d_{m}}
\;,
\end{equation}
the state space is two-dimensional and spanned by $U_{0}$ and
$\lambda$. In this state space, RKG determined the melting line, that
is, the function $U_{0}^{M}(\lambda)$. They introduced an effective
temperature $\tilde{T}$ ($kT$ in units of the Einstein phonon energy)
which is related to $U_{0}$ through
\begin{equation}
  \label{eq:11aa}
\tilde{T} = \Big[\frac{2}{3} \lambda^{2} f(\lambda) \beta U_{0} \Big]^{-1}  
\end{equation}
with a function $f(\lambda)$ given in Tab.~I of \cite{rkg}. For this
effective temperature, the RKG melting line is given by the function
$\tilde{T}^{M}(\lambda) = 0.00246 + 0.000274 \lambda$; it is plotted
in Fig.~(\ref{fig7}.a) as the thick solid line.  By identifying the
parameters in eq.~(\ref{eq:10}) with $U_{0}$ and $\lambda$ from
eq.~(\ref{eq:11}), i.e. by setting
\begin{equation}
  \label{eq:11a}
\beta U_{0} = \Big[ \frac{Z_{\eff} e^{\kappa_{\eff}a}}{1 + \kappa_{\eff}a}
\Big]^{2} \frac{\lambda_{B}}{d_{m}}; \quad\quad \lambda = \kappa_{\eff} d_{m}
\end{equation}
one can use this melting line also for colloidal systems. In the
$(Z_{\eff}, \kappa_{\eff}a)$ plane the melting line
$U_{0}^{M}(\lambda)$ becomes
\begin{equation}
  \label{eq:12}
Z_{\eff}^{M}(\kappa_{\eff} a) = (1 + \kappa_{\eff}a)e^{-\kappa_{\eff}a}\Big[
\beta U_{0}^{M}(\kappa_{\eff}a d_{m}/a) d_{m}/\lambda_{B}
\Big]^{1/2}
\end{equation}
with the function $U_{0}^{M}(\lambda)$ following from
$\tilde{T}^{M}(\lambda)$.  This line is plotted in Fig.~(\ref{fig7}.b)
for $\lambda_{B}/a = 0.012$ and four different values of $d_{m}/a = (4
\pi/3 \eta)^{1/3}$.  We recognize that a colloidal crystal melts if
either the salt concentration in the system or the colloidal charge or
volume-fraction is reduced.  One should realize that in general a
colloidal system has a four-dimensional state-space spanned by the
salt concentration, the colloidal charge, the volume fraction and
$\lambda_{B}/a$, while the phase-behavior of a pure Yukawa system is
determined just by $U_{0}$ and $\lambda$. That a Yukawa system is
capable of representing a colloidal system is simply due to the
special relation between the suspension parameters and the parameters
of the Yukawa potential expressed by eq.~(\ref{eq:13}). One has to
keep in mind however that the DLVO pair-potential in
eq.~(\ref{eq:10}), provides a reliable description only at low volume
fraction, but becomes inadequate at high volume fractions where
many-body interactions start playing a role. Then representing the
four-dimensional state-space of the suspension by the two dimensional
Yukawa state space ceases to work. The phase-boundary for $\eta=0.005$
in Fig.~(\ref{fig7}.b) is certainly a better estimate of the true
melting line at that volume fraction than it is the line for, say,
$\eta=0.05$. When plotted in Fig.~(\ref{fig7}.a), however, both curves
appear collapsed into a single one, namely the RKG curve.  In other
words, an experimental confirmation of the RKG line in the plot of
Fig.~(\ref{fig7}.a) at low volume-fractions does not guarantee a good
agreement between theory and experiment at large volume-fractions,
even though the data might cover the whole range of $\lambda$ in
Fig.~(\ref{fig7}.a).

Even when the representation in terms of a Yukawa liquid is taken for
granted, certain regions in the Yukawa phase-diagram of
Fig.~(\ref{fig7}.a) can never be explored by a colloidal suspension,
since the effective charge saturates at some finite value according to
Fig.~(\ref{fig5}.a). To identify the accessible region, we estimate
the saturated value of the effective colloidal charge by a formula
given by Oshima et al.  \cite{oshima82,trizac02}
\begin{equation}
  \label{eq:13}
Z_{\eff}^{\mathrm{sat}} = \frac{8 a}{\lambda_{B}} \frac{(1 +
  \kappa_{\eff} a)^{2}}{1 + 2 \kappa_{\eff} a}  \:.
\end{equation}
This expression is valid only in the limit $\eta \to 0$ which however
should be a reasonable approximation for the value of $\eta$ considered
here \cite{trizac02}. Inserting eq.~(\ref{eq:13}) into
eq.~(\ref{eq:11a}), transforming it to $\tilde{T}$ with
eq.~(\ref{eq:11aa}) and plotting it as a function of $\kappa_{\eff} a$
for fixed values of $\eta=0.03$ and $\lambda_{B}/a = 0.012$, we obtain
the thin solid line in Fig.~(\ref{fig7}.a): the region to the right of
this line cannot be reached with colloidal suspensions at that
combination of $\lambda_{B}/a$ and $\eta$ because $Z_{\eff}$ cannot
exceed $Z_{\eff}^{\mathrm{sat}}$.

Meijer and Frenkel \cite{frenkel} have located the melting line by
numerically determining the free energy of the fluid and the solid
phase (dot-dashed line in Fig.~(\ref{fig7}.a)).  For state points on
their melting line, they evaluated the rms values which are written
below the arrows in Fig.~(\ref{fig7}.a).  For small values of
$\lambda$ (soft potentials $\lambda<5$) the value was universal and
equal to 0.19 $d_{m}$, the very same value as used in \cite{rkg}; for
harder potentials (more salt, larger $\lambda$) they found
non-universal behavior and deviations of the rms value down to 0.15
$d_{m}$. While phase boundaries can be determined exactly only by
free-energy calculations, the latter are still rather time-consuming,
and a numerical determination of the rms is much easier to realize.
Since, the difference between the Frenkel/Meijer and the RKG line is
quite tolerable, compared to the magnitude of the many-body effects we
set out to investigate, we followed RKG and used the Lindemann
criterion in our study.

\section{Melting behavior of colloidal crystals in a many-body
  description}
\label{sec6}

Using the Poisson-Boltzmann-Brownian-dynamics (PB-BD) simulation
method described in sec.~\ref{sec3}, we have determined the melting
line in the phase-diagram of charge-stabilized colloidal suspensions
using Lindemann's rule. The phase-diagram in the $(Z,\kappa a)$-plane
shown in Fig.~(\ref{fig8}.a) represents our main result. It describes
the solid-liquid phase behavior of colloidal suspensions with all
many-body interactions among the colloids included. We use the
effective force parameters discussed and presented in sec.~\ref{sec4}
to plot the data of Fig.~(\ref{fig8}.a) in the
$(Z_{\eff},\kappa_{\eff} a)$ plane. This is done in
Fig.~(\ref{fig8}.b). In this plane, we can compare our melting line
with the predictions based on the established description in terms of
pairwise additive Yukawa potentials, i.e., with eq.~(\ref{eq:12})
(thick solid line in Fig.~(\ref{fig8}.b)).  Equivalently, we can
compare our data in the $(\tilde{T}, \lambda)$-plane used in the work
of RKG, by transforming the data of Fig.~(\ref{fig8}.b) with the help
of eq.~(\ref{eq:11a}) and (\ref{eq:11aa}). The resulting phase-diagram
is shown in Fig.~(\ref{fig9}).

As is evident from both plots, Fig.~(\ref{fig8}.b) and
Fig.~(\ref{fig9}), good agreement with RKG is obtained in our
high-salt calculation $\kappa a = 1.75$, corresponding to large values
of $\lambda$. This may be anticipated from the discussion in
sec.~\ref{sec4}. There, we obtained the effective parameters,
$Z_{\eff}$ and $\kappa_{\eff}$, for different state-points $\vec{X}$
by fitting our effective force curves to eq.~(\ref{eq:7}).  This was
done over the whole range of distances for the curves at high salt
concentration, but only over a limited range at short distances for
the low salt calculation because of the ''cut-off'' feature of the
force curves.  By this procedure we have mapped our original
non-linear many-body system onto an effective linear pairwise additive
Yukawa system. All the nonlinear and many-body effects are then folded
into an effective pair-potential and thus packed into the functions
$Z_{\eff}(\vec{X})$ and $\kappa_{\eff}(\vec{X})$.  We then need not
consider the system with all many-body interactions but can replace it
by a representative one-component model interacting via
eq.~(\ref{eq:7}) with $Z_{\eff}(\vec{X})$ and
$\kappa_{\eff}(\vec{X})$.  Since we know (see Fig.~(\ref{fig3}),
(\ref{fig4}) and (\ref{fig5})) that under high-salt conditions the
effective interactions show indeed a perfect and
configuration-independent Yukawa-like behavior, we must find back to
the RKG result when using the same procedure as RKG to determine the
melting line.  That our data point for $\kappa a =1.75$ lies again on
the RKG line is thus no surprise but proves the consistency of our
calculations.  However, this folding of many-body interactions into
effective Yukawa pair-potentials works only as long as i) the
interaction is not configuration dependent and ii) the effective
potential is adequately described by the functional form of the Yukawa
interaction which we know from sec.~\ref{sec4} is both not the case at
low salt.  And, indeed, one observes in Fig.~(\ref{fig8}.b) and
Fig.~(\ref{fig9}) pronounced deviations from the RKG line occurring in
the low salt regime (low $\kappa_{\eff}$ or $\lambda$), i.e., just in
the regime where we know that the many-body interactions are only
partly incorporated in the functions $Z_{\eff}(\vec{X})$ and
$\kappa_{\eff}(\vec{X})$. In other words, this deviation can have no
other reason than the configuration dependence of the interaction and
the departure of the effective pair forces from the Yukawa-pair force
at large distances observed in Fig.~(\ref{fig3}).  Taken together, all
these observations suggest what is essentially the main message of
this paper: that at high volume fraction and low-salt concentration a
charge-stabilized colloidal suspension is not adequately described by
the DLVO pair-potential in eq.~(\ref{eq:10}), and that as a
consequence of this fact, the melting line of a low-salt colloidal
system deviates from the RKG melting line of a simple Yukawa fluid.

Obviously, the comparison between our data and simulation results for
a system with strictly pair-wise additive potentials, like the RKG
data, is the only way to really demonstrate the influence of many-body
interactions. We stress that due to the parameters appearing in the
DLVO pair-potential in eq.~(\ref{eq:10}) this comparison is possible
only in the $(Z_{\eff},\kappa_{\eff} a)$ (or $(\tilde{T},\lambda)$)
-plane but not in the $(Z,\kappa a)$ plane. This explains why it was
necessary to calculate effective force curves as done in sec.
~\ref{sec3}.  We furthermore observe that all our data in
Fig.~(\ref{fig9}) lie on the left hand side of the curve labeled
$Z_{\eff}^{\mathrm{sat}}$ which we have introduced in
Fig.~(\ref{fig7}.a). The data point for our highest salt calculation
lies directly on the RKG line in Fig.~(\ref{fig9}), but also very near
to the $Z_{\eff}^{\mathrm{sat}}$ curve. This means that we have
explored with our simulations the maximum possible range of values for
$\lambda$; carrying out simulations for salt concentrations higher
than the highest considered here ($\kappa a = 1.75$) does not provide
more information as it always leads us back to the liquid state.

We should mention a word of caution regarding the Lindemann criterion.
The regime where we find deviations from the RKG line is just the
regime of soft potentials where the universal Lindemann value $rms =
0.19$ has been demonstrated to be good, see Fig.~(\ref{fig7}.a). We
therefore believe that using this value is perfectly justified for
values of $\lambda$ below 5. For high values of $\lambda$ it would be
more accurate to use a non-universal value corresponding to the work
of Meijer and Frenkel \cite{frenkel}. Then our points would not
approach the RKG line but the Meijer-Frenkel line, indeed, a rather
small quantitative shift, as Fig.~(\ref{fig9}) demonstrates. Since our
main interest here lies in the soft potential regime we ignored this
and used the universal value 0.19 throughout, just as in \cite{rkg}.
While for high salt concentrations our melting line is therefore
shifted up from the real melting line, we emphasize that for low salt
concentrations where we observe many-body effects the position of our
melting line should be correct.

If, as claimed above, the difference between our melting line and that
of RKG originates in the fact that the effective pair-interaction does
not have the functional form of a Yukawa potential, then one may
expect better agreement when comparing our data to a one-component
system with pair-interactions that are more adapted to the effective
force curves calculated in sec.~\ref{sec4}.  Guided by the cut-off
behavior observed in our effective-force curves at low salt
concentration as well as in the experiment \cite{klein,brunner}, we
suggest to model effective pair-interactions in colloidal systems by
truncated Yukawa interactions,
\begin{equation}
\label{eq:14}
f(r) = a\beta {\cal F}(r) = \left\{ \begin{array}{ll}
\frac{a\beta U_{0}}{d_{m}} 
\frac{1+\lambda r/d_{m}}{(r/d_{\nn})^{2}} e^{-\lambda r/d_{\nn}}& r\le r_{c}\\
0 & r > r_{c} \end{array} \right. \:,
\end{equation}
where the first line is just the derivative of eq.~(\ref{eq:11}). The
crucial point is that the cut-off $r_{c}$ in eq.~(\ref{eq:14}) is
chosen to be a multiple $x$ of the mean-distance, $r_{c} = x d_{m}$.
Through $r_{c}$ our interaction is then density-dependent.  In
eq.~(\ref{eq:14}), the Yukawa part results from microion-screening
while the hard cut-off is meant to represent the macroion screening.
If enough salt is present in the system, the cut-off is irrelevant, as
the Yukawa part has dropped to zero at $r=r_{c}$ anyway.

Using this model interaction, we carried out one-component MD
simulations and determined the solid-liquid phase-boundary again with
the Lindemann criterion, computing the rms displacement for various
combinations of $U_{0}$ and $\lambda$. Since the Yukawa system may
exist in two crystalline states with either FCC or BCC symmetry
\cite{rkg}, these simulations are carried out starting from FCC and
BCC crystals with 500 respectively 432 particles in a simulation box
with periodic boundary conditions.  For the FCC crystal,
Fig.~(\ref{fig10}.a), we have chosen a cutoff halfway between the
first and the second neighbor shell ($x=1.35$), the second and the
third ($x=1.5$), and directly before the third shell ($x=1.77$).  For
the BCC crystal, Fig.~(\ref{fig10}.b), we have chosen $x$ to be 1.5
(halfway between second and third shell), 1.7 (directly before third
shell) and 3.07.  With $x = 3.07 $ in the FCC crystal which is the
cut-off used in \cite{rkg} for technical reasons, the RKG melting line
was reproduced.  Upon decreasing the cut-off, we observe a systematic
shift of the melting line, both for the FCC and the BCC crystal,
occurring first at small values of $\lambda$ and becoming larger for
decreasing $r_{c}$.

The physical explanation for this shift is that a crystal looses
stability and thus melts earlier if one reduces the number of
repulsive bonds of a particle to its neighbors.  Here the cut-off in
the pair-interaction has been introduced to simulate the effect of the
macroion screening, and it is thus only natural to choose a cut-off
somewhere after the first neighbor-shell to compare it to our data in
Fig.~(\ref{fig9}). This is done in Fig.~(\ref{fig11}) where the dashed
lines are the melting lines for $x = 1.35$ (halfway between first and
second shells, 12 interacting particles) in a FCC configuration, and
$x = 1.5 $ (second shell, 14 interacting particles) in a BCC
configuration; the number of neighbors that are included in the
interaction is comparable in both cases.  Also given is the fcc-bcc
line determined in \cite{rkg}. The PB-BD melting points calculated
with effective force parameters $\kappa_{\eff}, Z_{\eff}$ from the
appropriate crystalline configuration agree remarkably well with these
two lines. This confirms the presumption made earlier that the
difference between the PB-BD points and the RKG line in
Fig.~(\ref{fig9}) is caused by the fact that the functional form of
the effective pair-interaction deviates from the pure Yukawa form.
Since the cut-off in our model pair-interaction in eq.~(\ref{eq:14})
is related in a simple way to the macro-ion shielding effect, this
further corroborates our conclusions about the importance of many-body
interactions for the solid-liquid phase behavior of colloidal
suspensions.

Fig.~(\ref{fig12}) is a schematic and simplified reproduction of the
phase diagram in Fig.~(\ref{fig9}) and (\ref{fig11}), containing again
the RKG line, the Meijer-Frenkel line (thick dashed line) and the
FCC-liquid and BCC-liquid line, derived from our calculations at
$\eta=0.03$. There are five regions denoted by letters.  Region $A$ is
liquid in both cases (RKG and ours). Region $B$ is solid (BCC) in RKG,
while we find it to be liquid due to many-body forces. Region $C$ is
solid (FCC) in RKG, we find a BCC solid phase instead. Region $D$ is a
FCC solid phase in both cases. The subregion denoted by $D_{A}$
becomes liquid in the exact free energy calculation \cite{frenkel}.
The transition from region $C$ to $D$ is speculative in this diagram.
The exact position should be defined using other techniques.

The good agreement between the results of the full calculation and
those of the simulations with the truncated Yukawa potential, obtained
at a volume fraction of $0.03$, does not allow to make final
predictions how this model potential performs at other
volume-fractions. Still, it is tempting to explore the effect which
the truncation of the Yukawa interaction has at other colloid
volume-fractions. We recall that the RKG melting line as well as our
lines in the $(\tilde{T},\lambda)$-plane (dashed and solid lines in
Fig.~(\ref{fig11})) are applicable to all colloid densities because
$d_{m}$ is taken here as the unit length scale. To be able to derive a
phase-diagram in the $(\eta,\kappa a)$-plane, we assume that the
effective colloidal charge is always saturated and that Oshima's
formula, eq.~(\ref{eq:13}), is valid at all salt concentrations and
all volume-fractions. As already done in Fig.~(\ref{fig7}.a), we can
then draw the curve $Z_{\eff}^{\mathrm{sat}}$ in our phase-diagram.
This limiting line is shown in Fig.~(\ref{fig13}.a) for three
different values of $\eta$.  From the crossing points of these lines
with the calculated melting lines, one can compute the phase-diagram
in the $(\eta, \kappa a)$-plane, shown in Fig.~(\ref{fig13}.b) for the
RKG line (solid line), our FCC-fluid line (empty circles) and our
BCC-fluid line (filled squares). Also given are the experimental data
points of Monovouskas and Gast, measured for colloidal spheres of
radius $a=66.7$ nm in water, i.e. $\lambda_{B}/a = 0.011$ (empty
squares).

We, first of all, observe that all three calculated melting lines lead
to the same solid-fluid line in the $(\eta,\kappa a)$-plane as long as
$\kappa a > 1$. This can be considered as the high-salt regime in
which colloidal interactions are well described by simple Yukawa
potentials, as we have already pointed out above. Below $\kappa a =
1$, differences are observed. The truncated Yukawa model predicts the
formation of a BCC and FCC pocket which results from the fact that the
line $Z_{\eff}^{\mathrm{sat}}$ in Fig.~(\ref{fig13}.a) crosses the
BCC-fluid and FCC-fluid line twice. This would imply a reentrance
behavior, but also that below a certain salt concentration and
volume-fraction the formation of stable crystals becomes impossible.
The exact topology of the phase diagram depends of course on the choice
of cut-off distance $x$; the feature itself, however, remains whatever
value of $x$ is chosen.

Bocquet et al. \cite{trizac02} have estimated the dependence of the
effective saturated charge on $\eta$ and $\kappa a$ and found a good
agreement between their theory and the data of Monovouskas and Gast
using the RKG melting line; they concluded that this is due to the
volume-fraction dependence of the effective charge. Using Oshimas
formula which ignores the volume-fraction dependence of
$Z_{\eff}^{\mathrm{sat}}$, we observe in Fig.~(\ref{fig13}.b) an
equally good agreement between the experimental data and the theory at
high salt concentration ($\kappa a > 1.5$). However, we believe that
this agreement is fortuitous. According to \cite{frenkel}, the
Lindemann criterion becomes questionable for hard potentials, i.e., at
high values of $\kappa a$. Furthermore, for large values of $\kappa a$
the hard-core interaction, neglected in the RKG calculation, must be
taken into account. Thirdly, there is definitely a non-negligible
$\eta$-dependence of the saturated effective charge \cite{trizac02}.
And, finally, it is not clear whether the effective charge in the
experiment can at all be assumed to be saturated. The good agreement
in Fig.~(\ref{fig13}) at $\kappa a >1.5$ might thus result from a
happy cancellation of errors introduced by all these assumptions.

Oshima's formula, neglecting the contribution of the counterions to
$\kappa a$, becomes better at smaller $\eta$, where also RKG and the
Frenkel Meijer line agree; so the comparison made in
Fig.~(\ref{fig13}.b) is reasonable at $\kappa a \le 1.5$. The
agreement is fair in the range $0.8<\kappa<1.5$, but not below $\kappa
a =0.8$. Still, it is interesting to observe that the suspension in
the experiment ceases to crystallize at sufficiently low volume
fraction, a feature that cannot be understood in a description in
terms of pure Yukawa potentials predicting a solidification for all
volume-fractions.  The truncated Yukawa model potential, on the other
hand, predicts that solidification below a certain volume-fraction is
impossible.  While this qualitative conclusions can certainly be
drawn, the truncated Yukawa model is probably too approximate at low
volume-fractions for any quantitative comparison.

We close this section with a few further comments: (i) It should be
noted that beyond the melting line, in what is referred to as 'liquid'
phase in our phase-diagrams, there is, in reality, a region of
solid-liquid phase coexistence whose shape and extent remains to be
determined.  (ii) Studies treating colloidal suspensions as hard-core
Yukawa fluids \cite{meijer97,azhar00} also revealed a phase-behavior
different from that of a pure Yukawa fluids. However, these
differences occur only in the limiting case of weakly charged colloids
or at extremely high salt concentration which is not considered here.
In our treatment, hard-core interactions are included, but play
virtually no role.  (iii) To investigate the effect that a possible
size polydispersity of the colloids may have on the melting line, we
have performed MC computer studies for particles interacting via $u(r)
= U_{0}^{\prime} e^{-\lambda r/d_{m}}/ (r/d_{m})$ where
$U_{0}^{\prime} = U_{0}(1+\gamma)$ with $\gamma$ a Gaussian random
number with mean value zero and variance $\sigma$. We found virtually
no effect as long as $\sigma \le 0.01$. Only when $\sigma$ was as high
as 0.1, a shift of the melting was seen which then however was as
strong as the effect observed here. (iv) In highly de-ionized systems
where colloids are dispersed in organic solvents, the inverse
screening length can become as large as a few microns
\cite{phillipse}. In a very recent experiment, Yethiraj and van
Blaaderen \cite{yethiraj} have studied the crystallization behavior of
$\eta=0.02$ charge-stabilized colloidal suspensions in an organic
solvent where the inverse screening length was 12 $\mu m$.  In these
systems, many-body effects should then be much more pronounced than in
the aqueous system considered here.

\section{Conclusion}
\label{sec7}

The following physical picture emerges from our study, and is
described here as a summary of our results. Three-body forces between
charged colloids are attractive. By performing effective force
calculations, all three- and higher-body interactions are folded into
an effective pair-potential; they are integrated out, similar to what
is suggested by eq.~(\ref{eq:1}). At high salt concentration, these
effective pair interactions are found to be Yukawa-like at all
distances. However, if the salt concentration is low ($d_{m} \kappa
\sim 1$), they are Yukawa-like only at short distances, and decay down
to zero much faster than a Yukawa potential, at larger distances ($r >
d_{m}$).  This is due to the folded-in many-body interactions, or,
expressed in another way, because the macroions surrounding a given
pair of interacting colloids have a screening effect just as the
microions do. The melting behavior of a crystal for colloids
interacting via such truncated Yukawa pair-interactions is different
from that of a pure Yukawa system. For a colloidal crystal whose
stability rests on repulsive pair-forces, taking away bonds by
truncating the potential has a destabilizing effect on the crystal: it
melts more easily (at lower effective temperature, or higher effective
charge) than it would if the Yukawa interactions were not truncated.

We have qualitatively confirmed this picture by effective force
calculations, showing explicitly the effect of other macroions on an
interacting pair of colloids, and by a full PB-BD simulation of
colloidal crystals, yielding a melting line which for the system
considered here ($\eta=0.03$, $\lambda_{B}/a = 0.012$) and at
low-added salt, shows marked deviations from the classic melting line
of a Yukawa system determined by RKG \cite{rkg}. Our melting line
could be well reproduced within a simple one-component model of
particles interacting via truncated Yukawa potentials. In essence, in
low-salt suspensions at high colloid volume fraction, i.e., when
many-body interactions start playing a role, we predict a liquid phase
where one finds a BCC solid in the pair-Yukawa description, and a BCC
solid where the Yukawa system is a FCC solid.

The important influence of many-body interactions on the
phase-behavior has long been recognized in rare-gas systems. But in
colloidal systems -- which are much better suited to study this
question, as the interactions are tunable -- this question has not
received much attention. Experimental studies of the melting /
crystallization behavior of charged colloids, under low salt
conditions and possibly also in organic solvents, are highly
desirable. Another interesting open question is the effect of
many-body interactions on the gas-liquid phase behavior.  For example,
the Axilrod-Teller interaction potential has been shown to have a
measurable impact on the gas-liquid equilibrium of argon
\cite{bukowski01}. The question of a possible gas-liquid phase
coexistence in colloidal systems is currently studied mainly with
volume term theories \cite{levin02}.  It would be interesting to see
the effect of three-body interactions on the gas-liquid
phase-coexistence in colloidal systems. A rough estimate of this
effect, within a simple van-der Waals picture, is given in
\cite{russ}, showing that, in principle, attractive three-body
interactions provide enough cohesive energy for a gas-liquid phase
coexistence to become possible. More definite answers to these
questions, can be expected either from studies using approaches
similar to our Poisson-Boltzmann-Brownian dynamics method or more
rigorous primitive model studies, similar to those presented by Linse
and Lobaskin \cite{linse4359,linse4208,linse3917}.


\newpage

{\bf Figure captions} \\

\ \\

{\bf fig1} \\
Dimensionless effective colloid-colloid force $f(r)$ between two out
of $N=108$ macroions arranged in a FCC configuration, as obtained from
our PB calculation (symbols) and from the Yukawa pair-force in
eq.~(\ref{eq:7}) (solid line) with effective force parameters determined 
from a fit to the data as $Z_{\eff}=1080$ and $\kappa_{\eff}=1.98$. 
Inset: force curve from the main figure multiplied by 
$(r/a)^{2}/(1+\kappa_{\eff} r)$ and plotted logarithmically so that a
Yukawa force appears as a straight line with slope $-\kappa_{\eff}$.  
The system parameters are: volume fraction $\eta=0.03$, Bjerrum length
$\lambda_{B} = 0.012 a$, bare charge $Z=3000$, and salt concentration
$\kappa a = 2.0$.\\

\ \\

\begin{figure}
\includegraphics[width=0.5\textwidth]{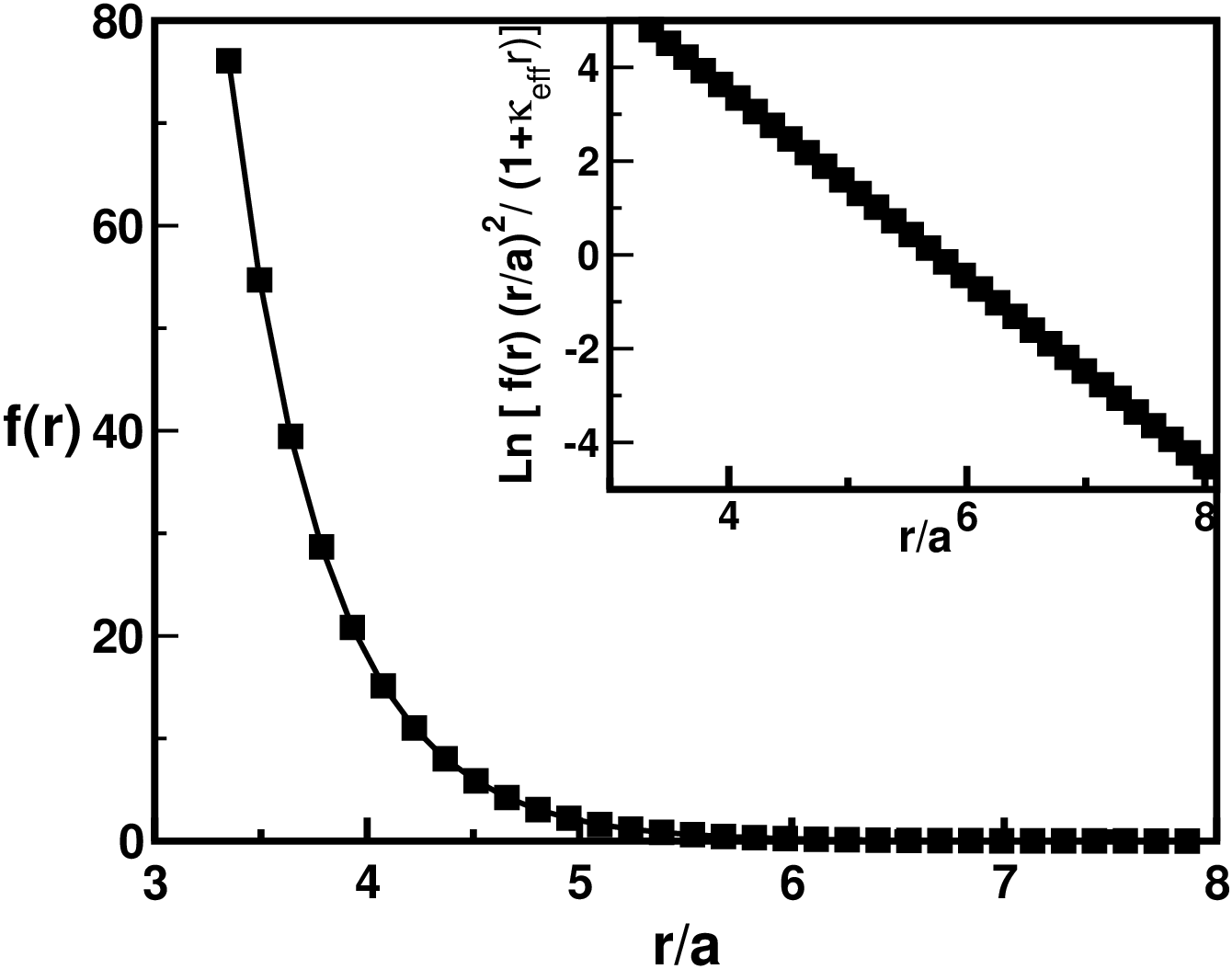}
\caption{\label{fig1}} 
\end{figure}

{\bf fig2} \\
Effective force curve like in Fig.~(\ref{fig1}), 
but for $Z=1000$ and $\kappa a=0.2$ (low salt condition), 
plotted as in the inset of Fig.~(\ref{fig1})
so that a Yukawa force would appear as a straight line.
The dot-dashed line is the best-fitting Yukawa interaction
at small particle separation which here is given 
in units of the mean-distance $d_{m}$.
The other parameters are the same as in Fig.~(\ref{fig1}).\\

\ \\

\begin{figure}
\includegraphics[width=0.5\textwidth]{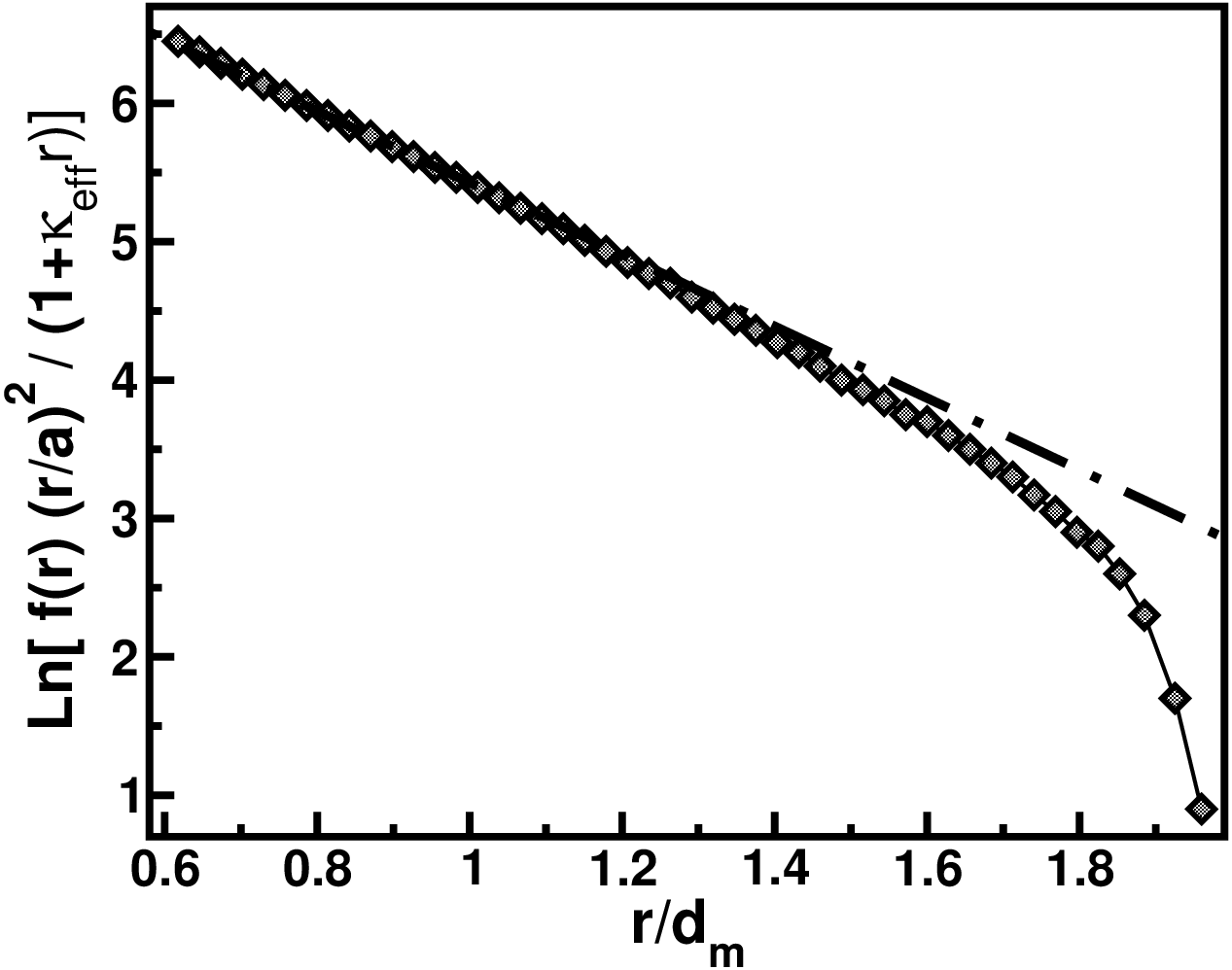}
\caption{\label{fig2}} 
\end{figure}

\begin{figure}
\includegraphics[width=0.5\textwidth]{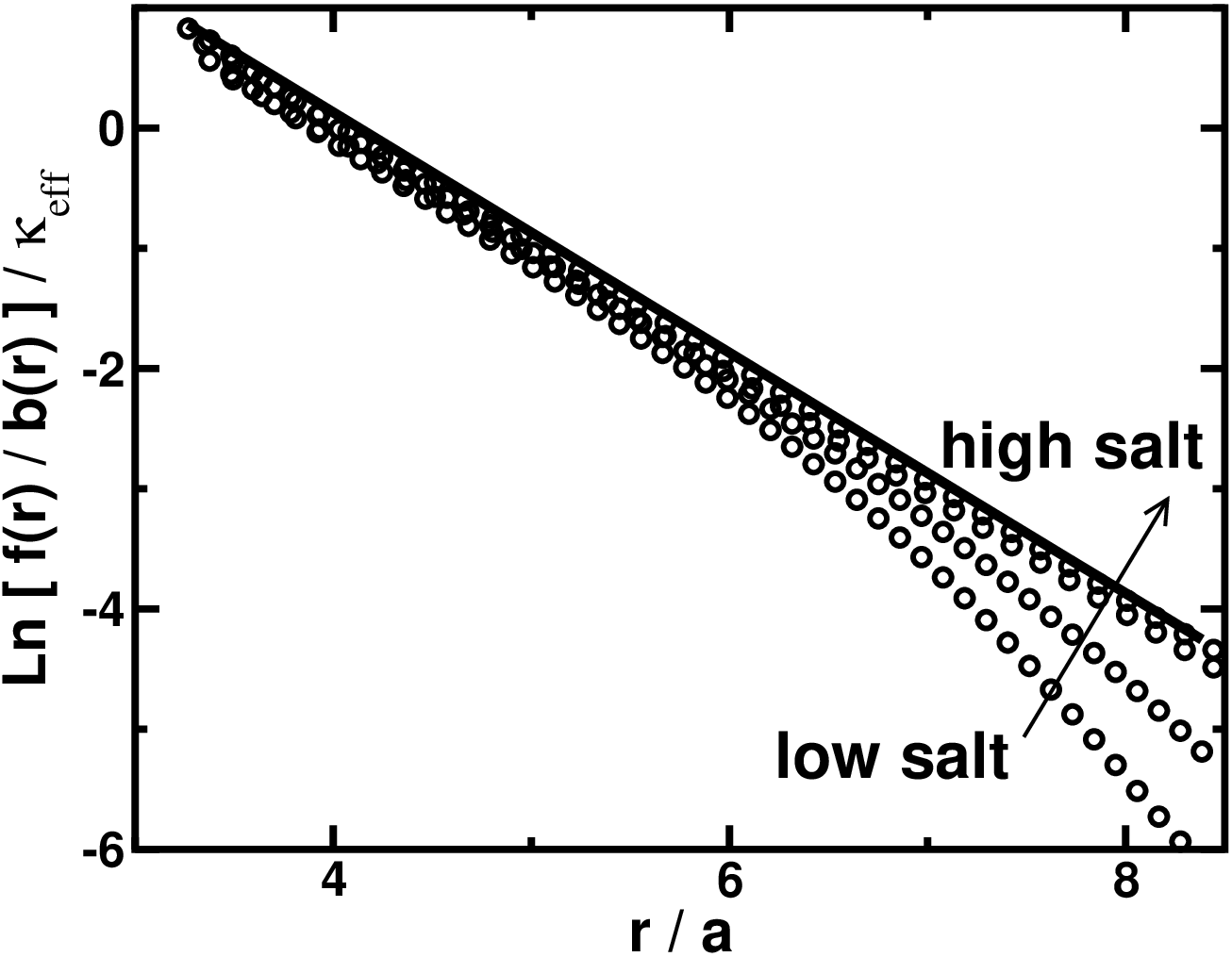}
\caption{\label{fig3}}
\end{figure}

\begin{figure}
\includegraphics[width=0.5\textwidth]{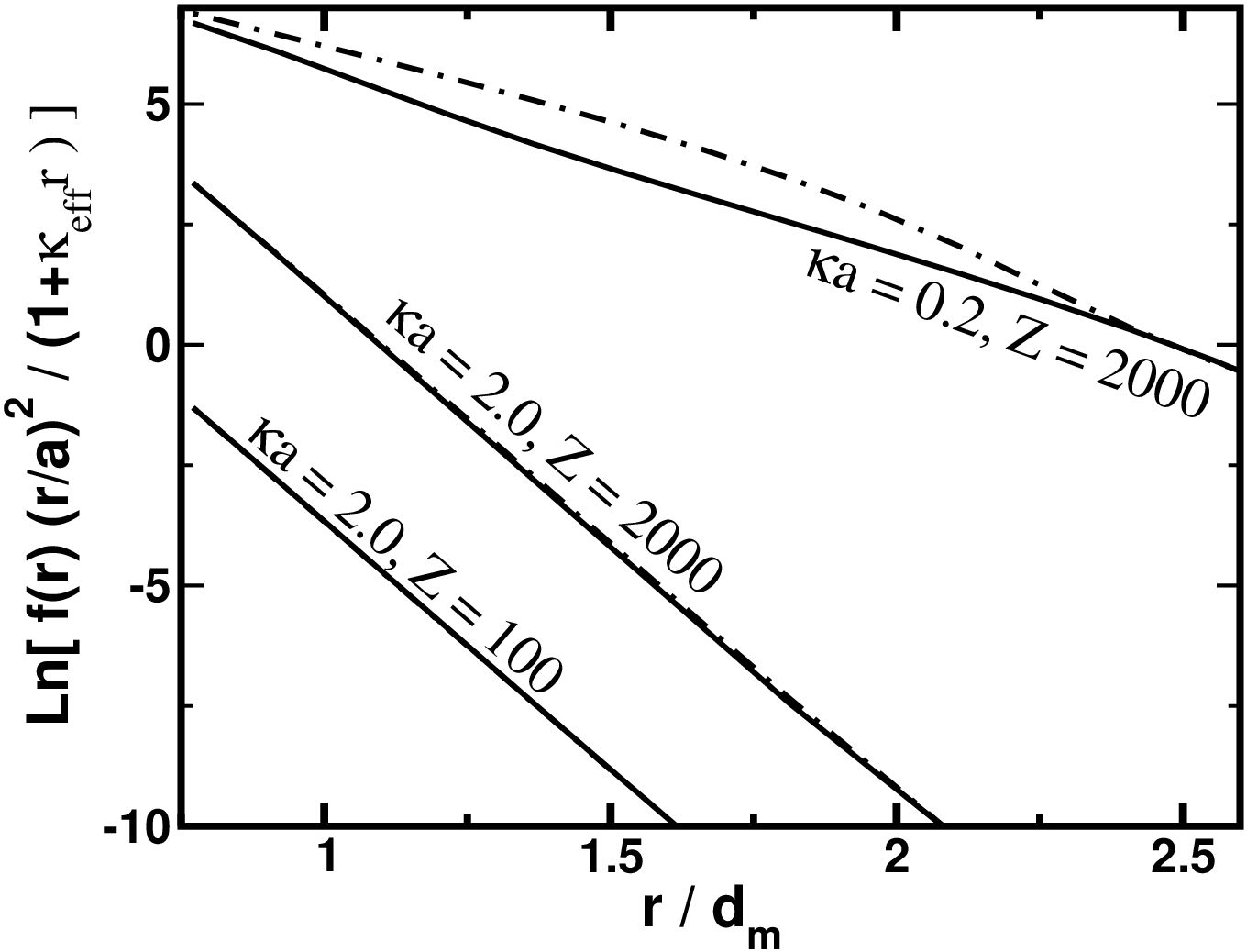}
\caption{\label{fig4}}
\end{figure}

{\bf fig3} \\
Effective force curves in a FCC configuration for different salt
concentrations ($\kappa a=0.1, 0.2, 1.0, 2.0$, from bottom to top).
The calculated forces $f(r)$ are plotted such that a Yukawa pair-force
following the DLVO theory, eq.~(\ref{eq:7}) appears as a straight line
with the slope -1 (solid line). Deviations from Yukawa-like behavior
become visible for $\kappa a < 1.0$.  Remaining parameters as in
Fig.~(\ref{fig2}).\\

\ \\

{\bf fig4} \\
Effective force curves, plotted like in the inset of
Fig.~(\ref{fig1}), for three different combinations of $Z$ and $\kappa
a$ as indicated. For each combination, both a FCC configuration
(dot-dashed lines) and a BCC configuration (solid lines) are analyzed.
For a short screening length, $\kappa a=2.0$, the pair of curves lie
on top of each other over a wide range of charges, $Z=100 \ldots
2000$.  When the screening length is increased, $\kappa a=0.2$,
however, there is a clear deviation between both.  $r$ is given here
in units of the mean-distance $d_{m}$.  The other parameters are again
the same as in Fig.~(\ref{fig1}).\\

\ \\

{\bf fig5} \\
Effective charge (a) and effective screening parameter (b) versus bare
charge, obtained from fitting Yukawa pair-forces, eq.~(\ref{eq:7}), to
effective force curves in BCC (filled squares) and FCC (empty circles)
configurations, for four salt concentrations as indicated and $\eta
=0.03$, $\lambda_{B}/a=0.012$.  The solid lines are the predictions of
the PB cell model \cite{alexander} for comparison. In (b), the curve
for $\kappa a = 0.5$ is omitted for clarity. Dotted lines are guide to
the eyes.\\

\ \\

\begin{figure*}
\includegraphics[width=0.45\textwidth]{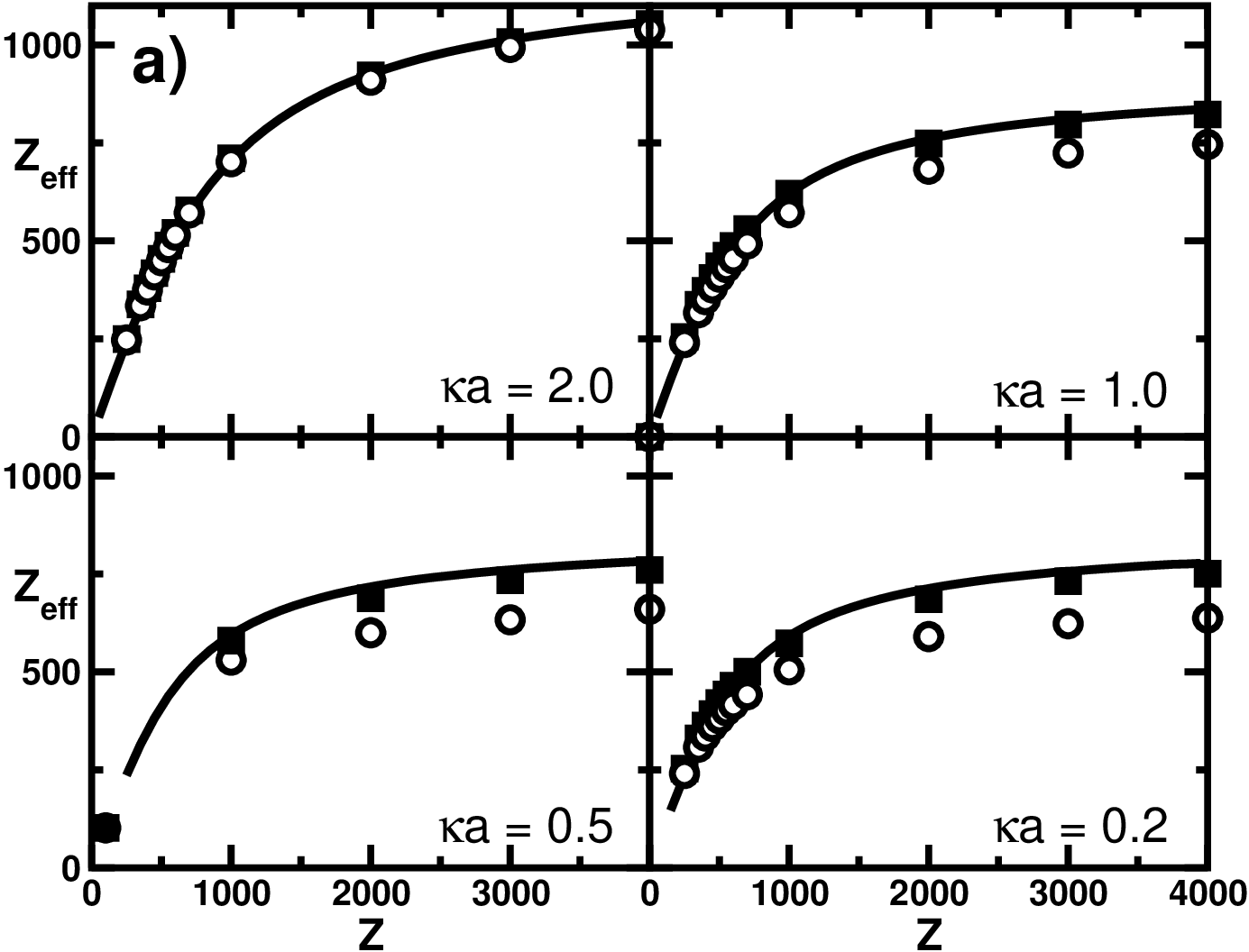}\hfill
\includegraphics[width=0.45\textwidth]{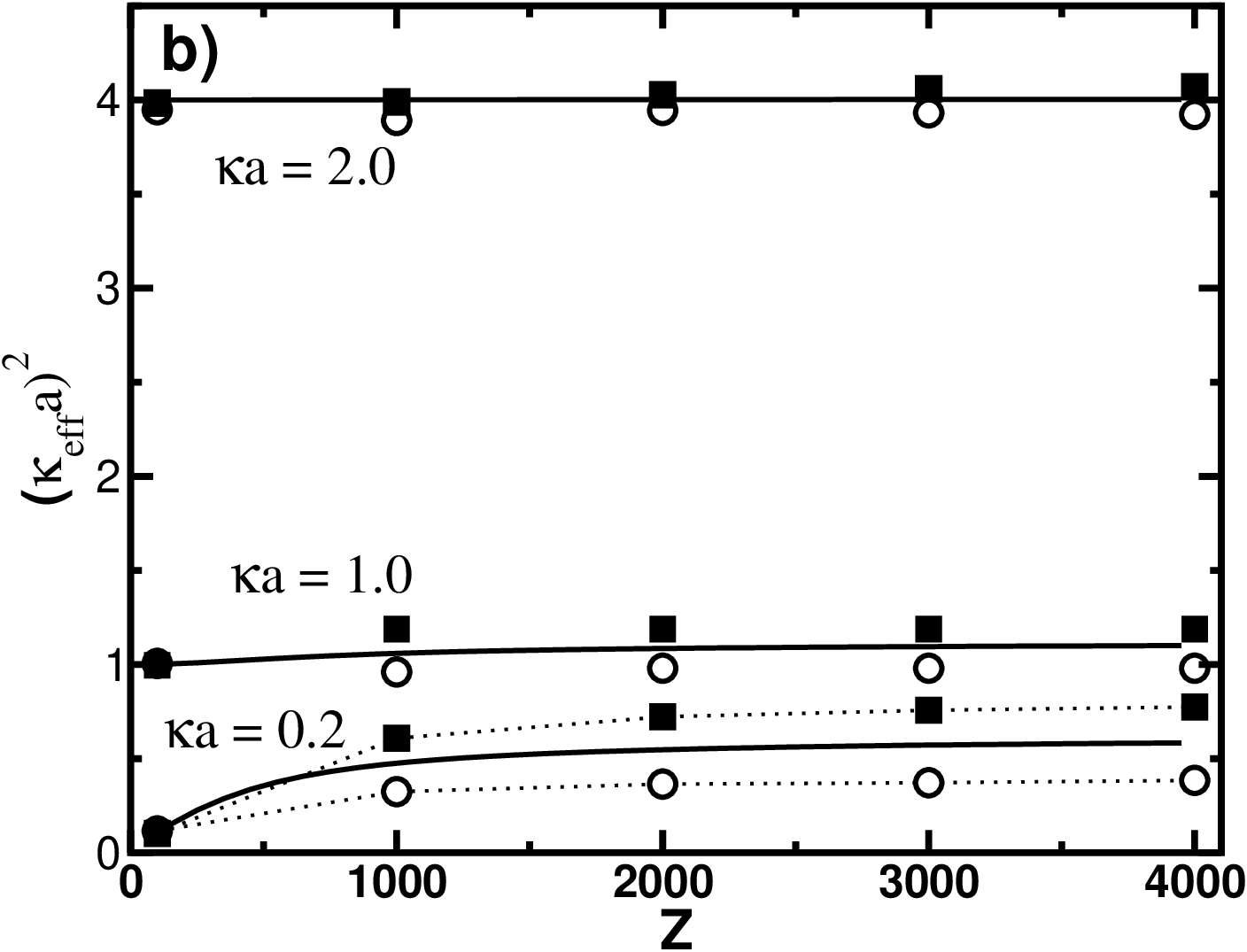}
\caption{\label{fig5}}
\end{figure*}

\begin{figure*}

\ \\

\includegraphics[width=0.45\textwidth]{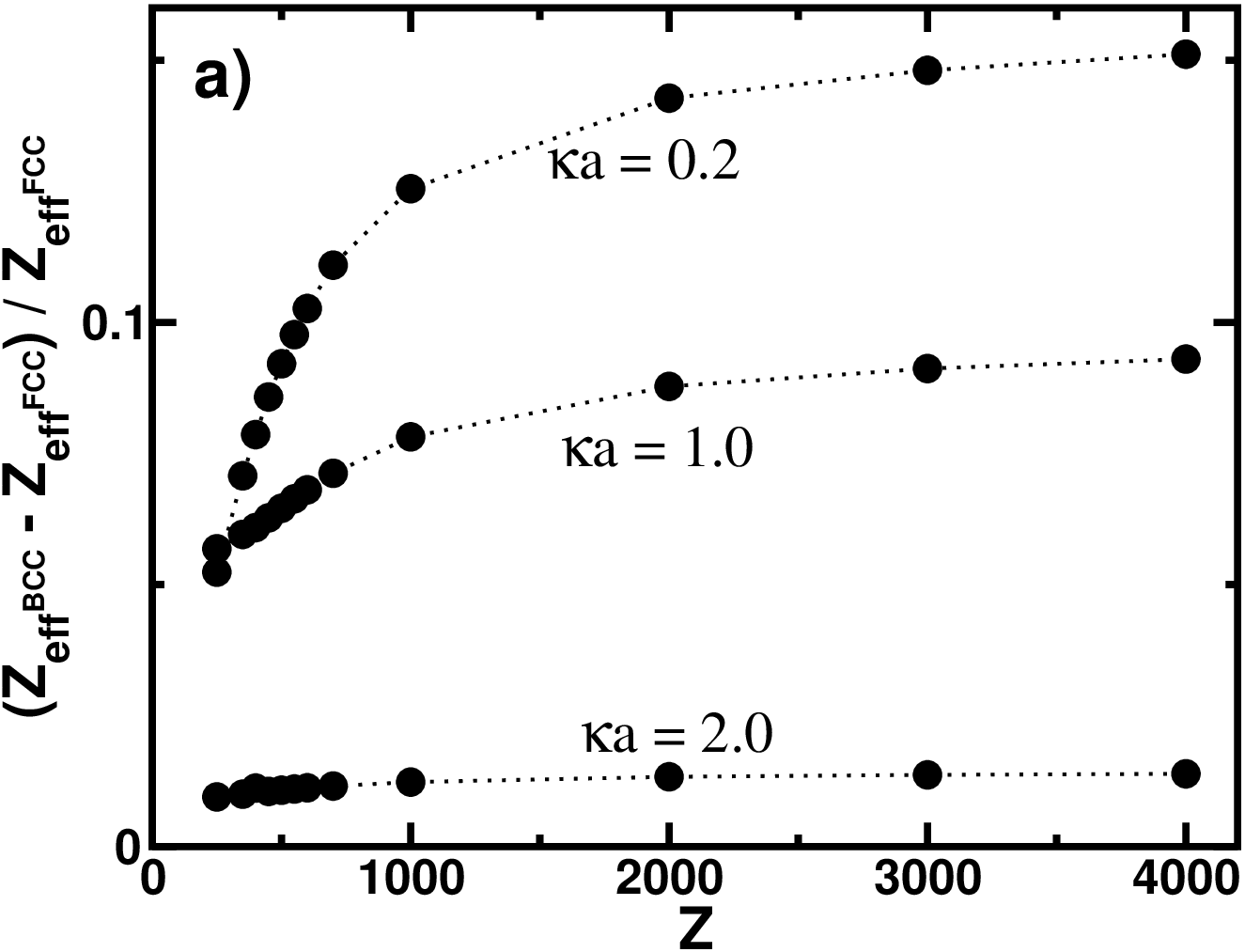}\hfill
\includegraphics[width=0.45\textwidth]{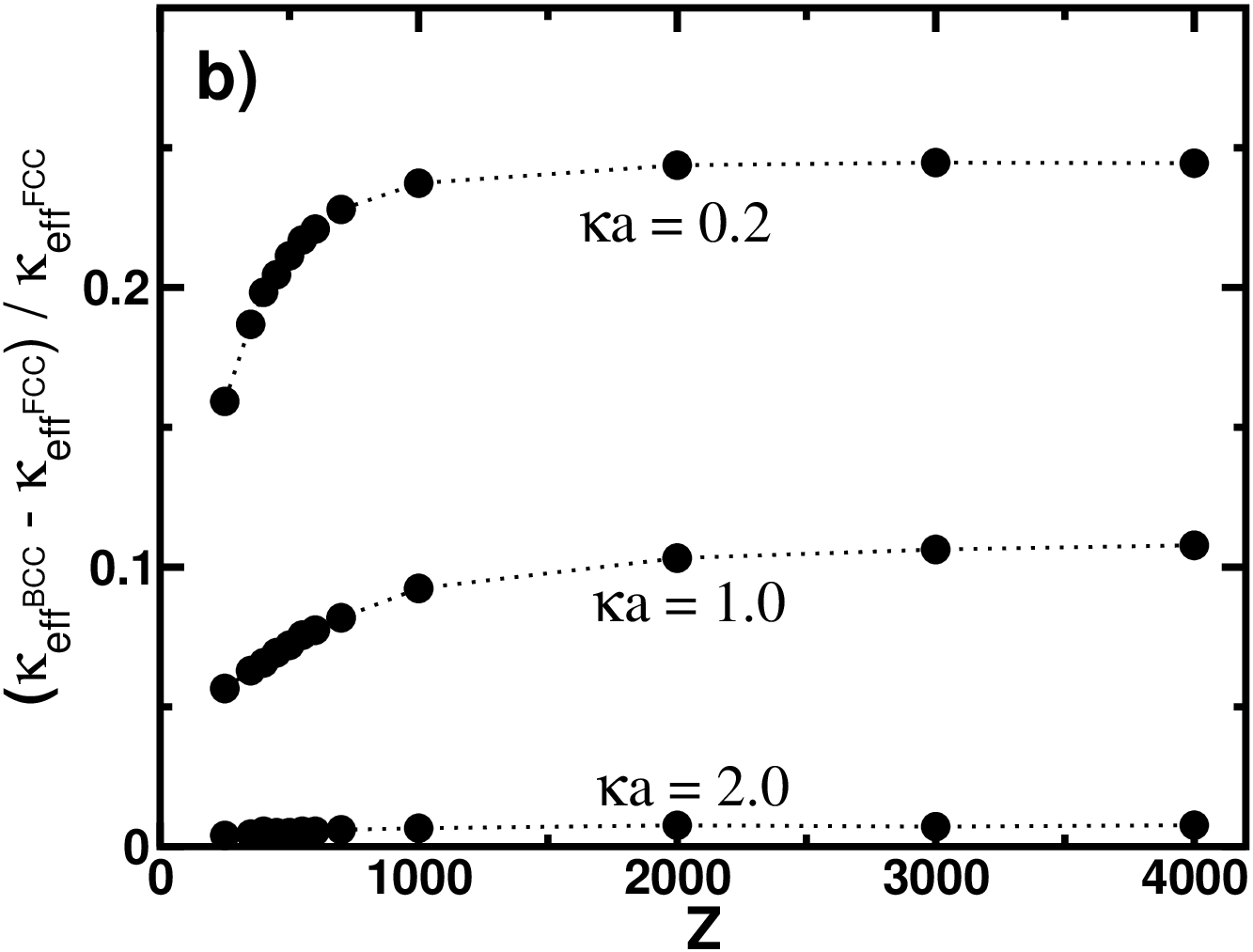}
\caption{\label{fig6} }
\end{figure*}

{\bf fig6} \\
Relative differences between the BCC- and FCC-based effective
Yukawa force parameters shown in Fig.~(\ref{fig5}).\\

\ \\

{\bf fig7} \\
(a) Melting line in the state space of a simple Yukawa fluid spanned
by the reduced temperature $\tilde{T}$ and the ratio of the
mean-distance to the screening length $\lambda = \kappa d_{m}$,
according to Robbins et al.  \cite{rkg} (thick solid line) and Meijer
and Frenkel \cite{frenkel} (dot-dashed line). The region to the right
of the thin solid line labeled by $Z_{\eff}^{sat}$ can not be explored
by a colloidal suspension at a volume fraction $\eta=0.03$ and
$\lambda_{B}/a = 0.012$, see text. Numbers below arrows indicate the
rms values for state-points on the Meijer-Frenkel line as determined
in \cite{frenkel}. For soft potentials there is a good agreement with
the Lindemann value of 0.19.  (b) Melting line of RKG transformed with
eq.~(\ref{eq:12}) into the state space spanned by $(Z_{\eff},
\kappa_{\eff} a)$ of a charge-stabilized colloidal suspension, for
various volume fractions $\eta$ ($\lambda_{B}/a = 0.012$). \\

\ \\ 

{\bf fig8} \\
(a) Solid-liquid phase diagram of a colloidal suspension, spanned by
the colloidal charge $Z$ and the salt concentration $\kappa a$ for a
volume fraction $\eta=0.03$ and $\lambda_{B}/a=0.012$.  Points on the
melting line are obtained by applying the Lindemann criterion to rms
data from a PB-BD simulation of a FCC crystal (empty circles) and a
BCC crystal (filled squares).  Many-body interactions between the
colloids are fully included in these simulations. The dashed line is a
guide to the eyes.  (b) Data from (a) transferred into the
$(Z_{\eff},\kappa_{\eff} a)$ plane by using the effective parameters
from Fig.~(\ref{fig5}).  For high-salt concentration, the simulation
data agree well with the RKG melting line of a Yukawa fluid as
obtained from eq.~(\ref{eq:12}) (thick solid line), while for low-salt
concentration there are pronounced differences due to the presence of
many-body interactions in the suspension.\\

\ \\

{\bf fig9} \\
Melting line of a simple fluid of particles interacting via truncated
Yukawa pair-forces, eq.~(\ref{eq:14}), as obtained by applying the
Lindemann criterion to rms data from a MD simulation of a FCC crystal
(empty circles in (a)) and a BCC crystal (filled squares in (b)).  The
Yukawa interaction is truncated at $r_{c} = x d_{m}$
($d_{m}=\rho^{-1/3}$). The value of $x$ is given as label to
the curves; $x=3.07$ is the value used by RKG (thick solid line).\\

\ \\ 

\begin{figure*}
\includegraphics[width=0.45\textwidth]{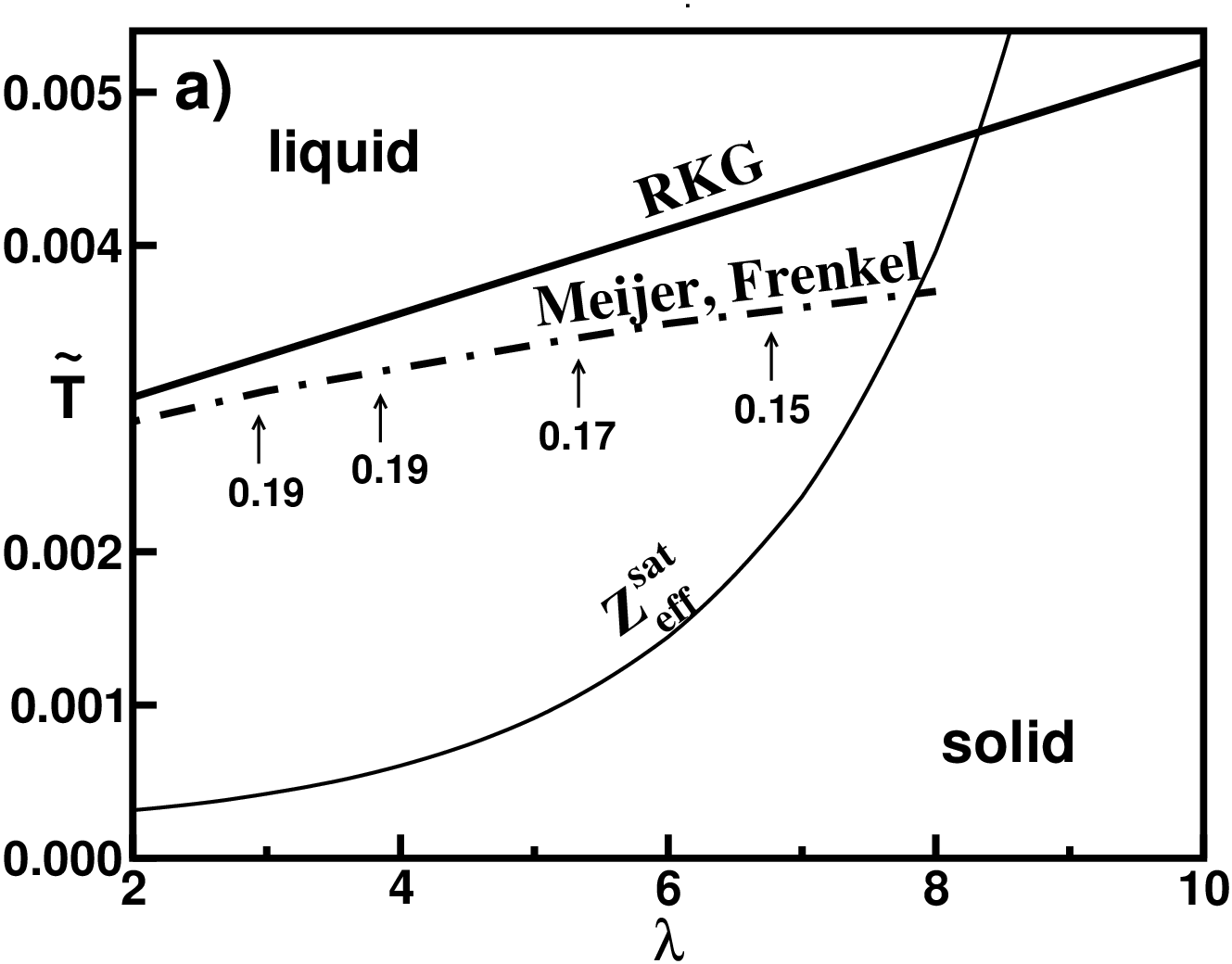}\hfill
\includegraphics[width=0.45\textwidth]{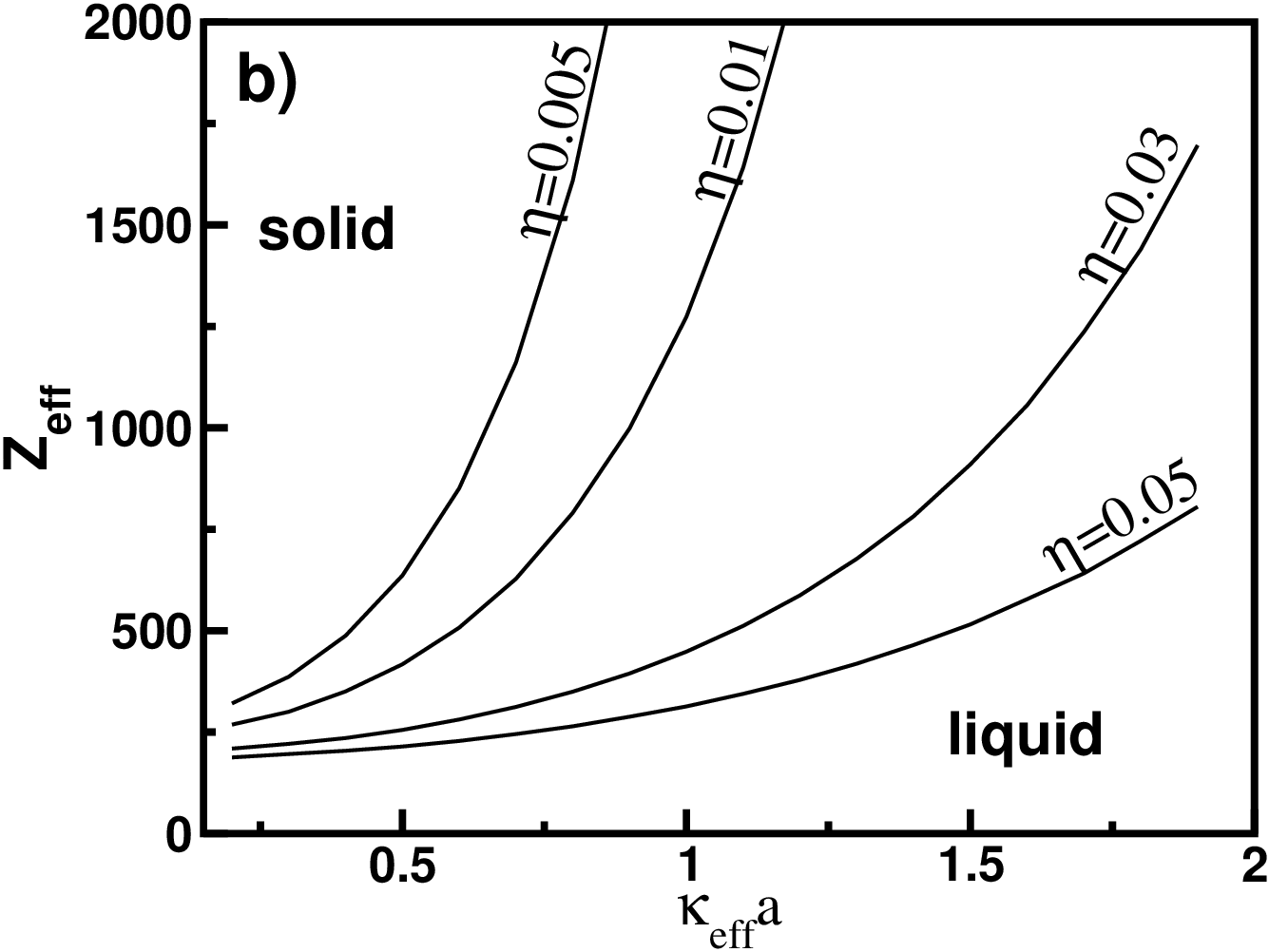}
\caption{\label{fig7} }
\end{figure*}

{\bf fig10} \\
Data from Fig.~(\ref{fig8}.b) transferred into the
$(\tilde{T},\lambda)$ plane used in the work of RKG ({\em cf.}
Fig.~(\ref{fig7}.a)).  The symbols for the PB-BD data are as defined
in Fig.~(\ref{fig8}.a), the meaning of the lines is explained in the
figure caption of
Fig.~(\ref{fig7}.a).\\

\ \\ 

{\bf fig11} \\
The $x=1.35$ FCC-liquid and the $x=1.5$ BBC-liquid line of
Fig.~(\ref{fig10}) (dashed curves) in comparison with the PB-BD
simulation results presented in Fig.~(\ref{fig9}). The solid lines
are the melting line and the bcc-fcc line determined by RKG.\\

\ \\

{\bf fig12} \\
Schematic reproduction of the phase diagram in Fig.~(\ref{fig9}) and
(\ref{fig11}), letters and hatched regions are explained in the
text.\\

\ \\ 

{\bf fig13} \\
(a) Phase-diagram of Fig.~(\ref{fig11}) together with the limiting
curve $Z_{\eff}^{sat}$ introduced in Fig.~(\ref{fig7}.a), at three
different volume fractions as indicated ($\lambda_{B}/a = 0.012$).
(b) Comparison between experimental and theoretical phase-diagrams of
a charge-stabilized colloidal suspension in the plane spanned by the
volume-fraction and the salt concentration, derived from the RKG
melting line (solid line), the FCC-fluid (empty circles) and BCC-fluid
(filled squares) line of the truncated Yukawa model system. The
experimental points are reproduced from Ref.~\cite{gast89}. The inset
shows the experimental points against the RKG melting line in a larger
range of $\kappa a$.

\newpage

\ \\

\newpage

\ \\

\newpage

\ \\

\newpage

\ \\

\newpage

\ \\

\newpage

\ \\

\newpage

\ \\

\newpage

\ \\

\begin{figure*}
  \includegraphics[width=0.45\textwidth]{fig8a}\hfill
  \includegraphics[width=0.45\textwidth]{fig8b}
\caption{\label{fig8} }
\end{figure*}

\newpage

\ \\ 

\begin{figure}
\includegraphics[width=0.45\textwidth]{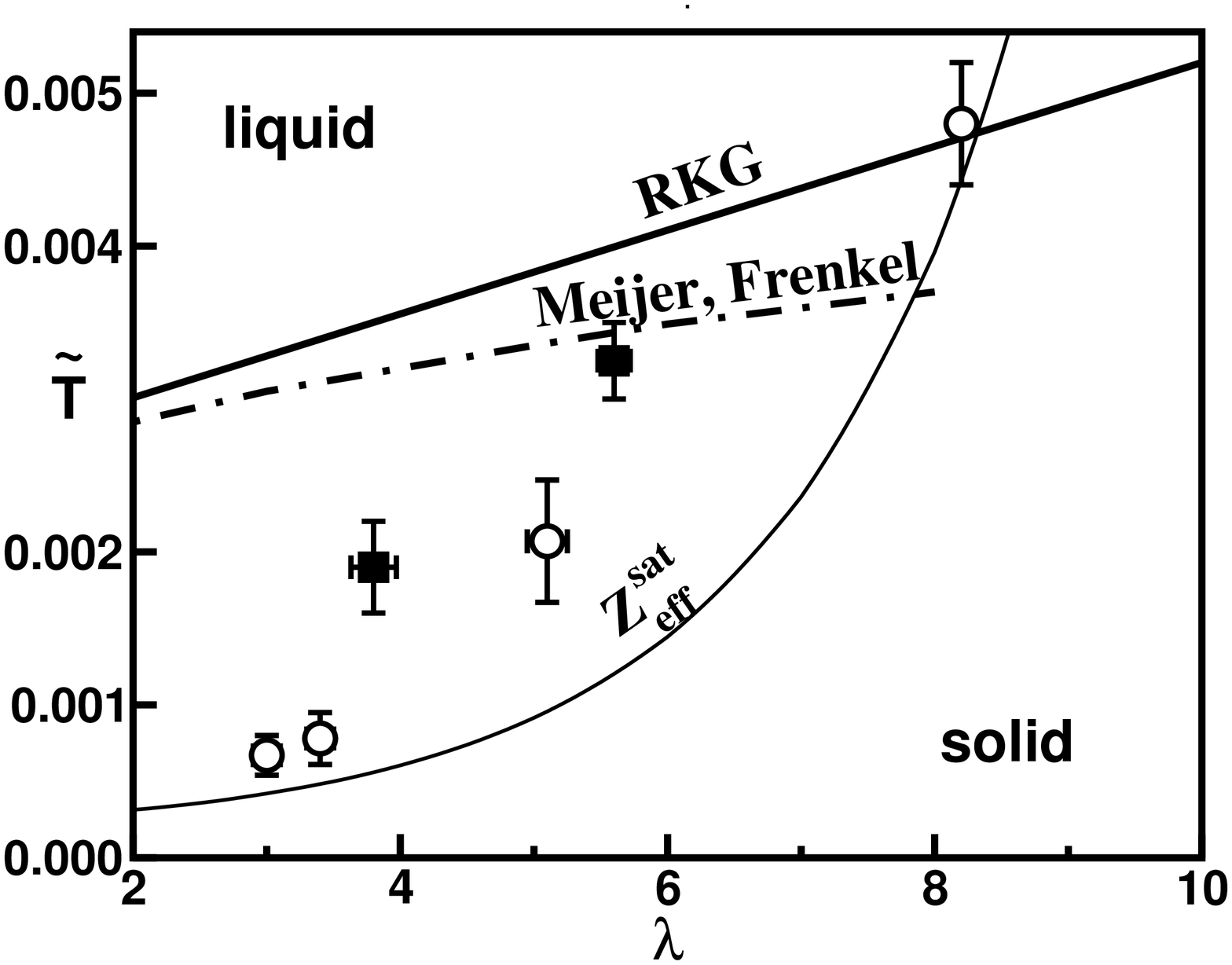}
\caption{\label{fig9} }
\end{figure}

\newpage

\ \\

\begin{figure*}
\includegraphics[width=0.45\textwidth]{fig10a}\hfill
\includegraphics[width=0.45\textwidth]{fig10b}
\caption{\label{fig10}} 
\end{figure*}

\newpage

\ \\

\begin{figure}
\vspace{1cm}
\includegraphics[width=0.45\textwidth]{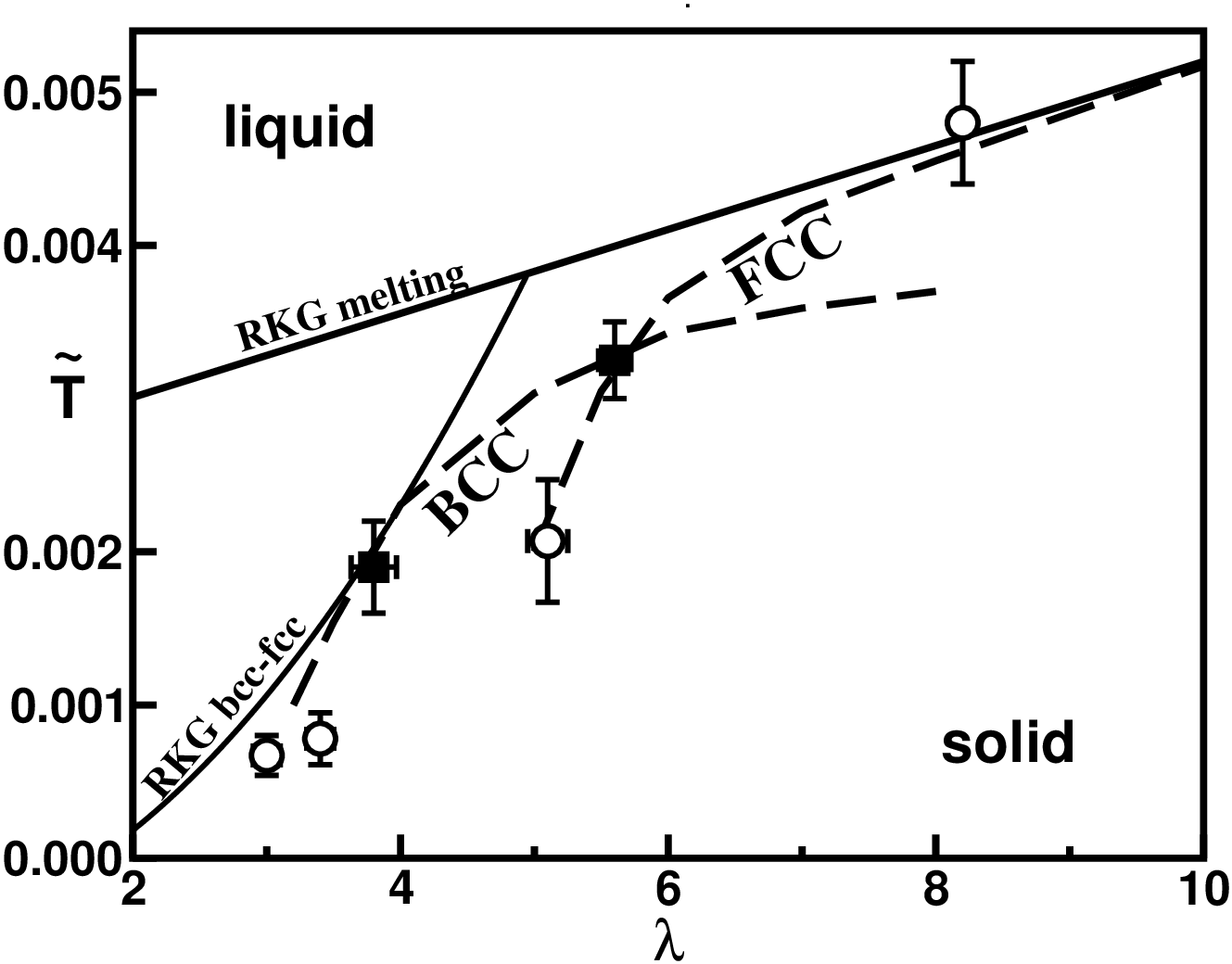}
\caption{\label{fig11} }
\end{figure}

\newpage

\ \\

\begin{figure}
\includegraphics[width=0.45\textwidth]{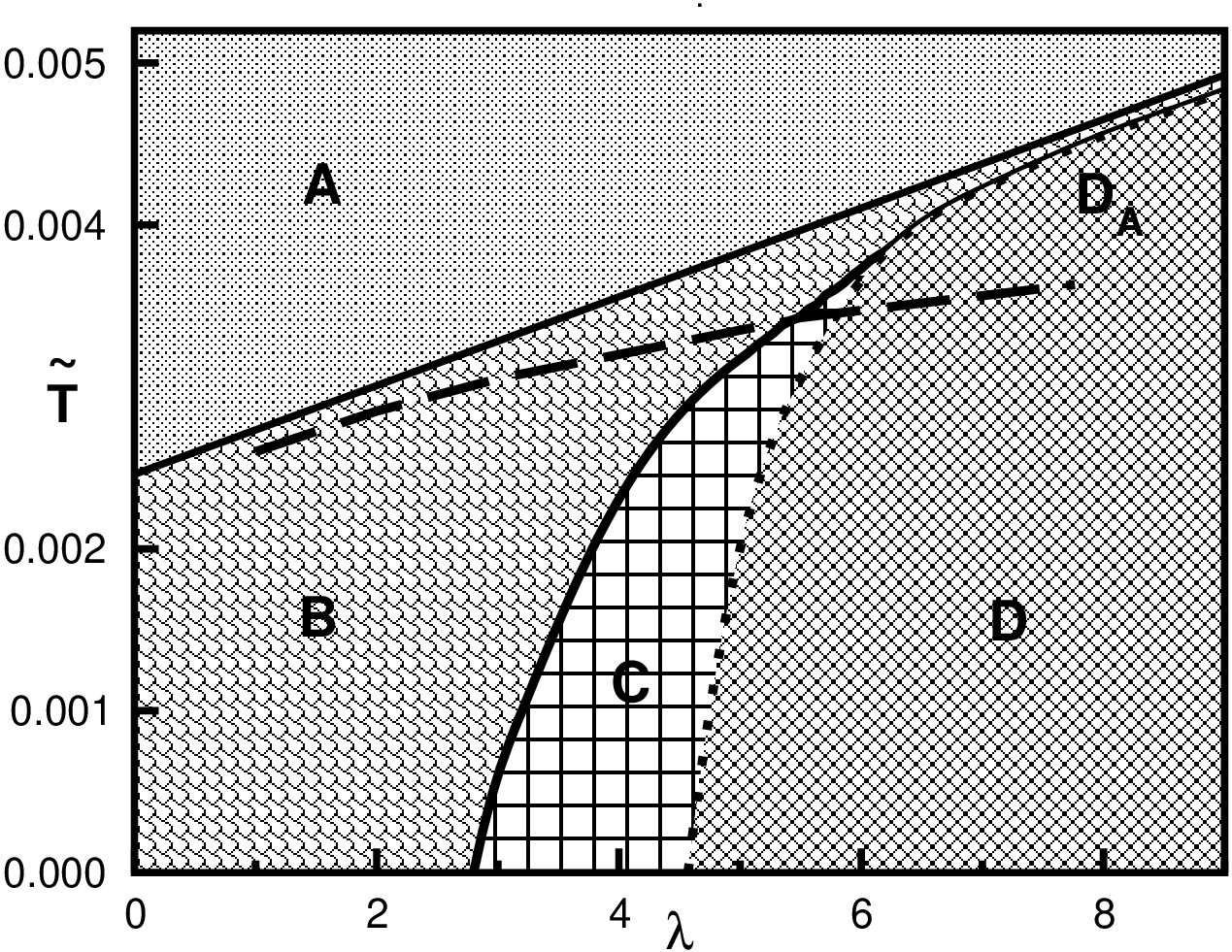}
\caption{\label{fig12} }
\end{figure}

\newpage

\ \\

\begin{figure}
\includegraphics[width=0.45\textwidth]{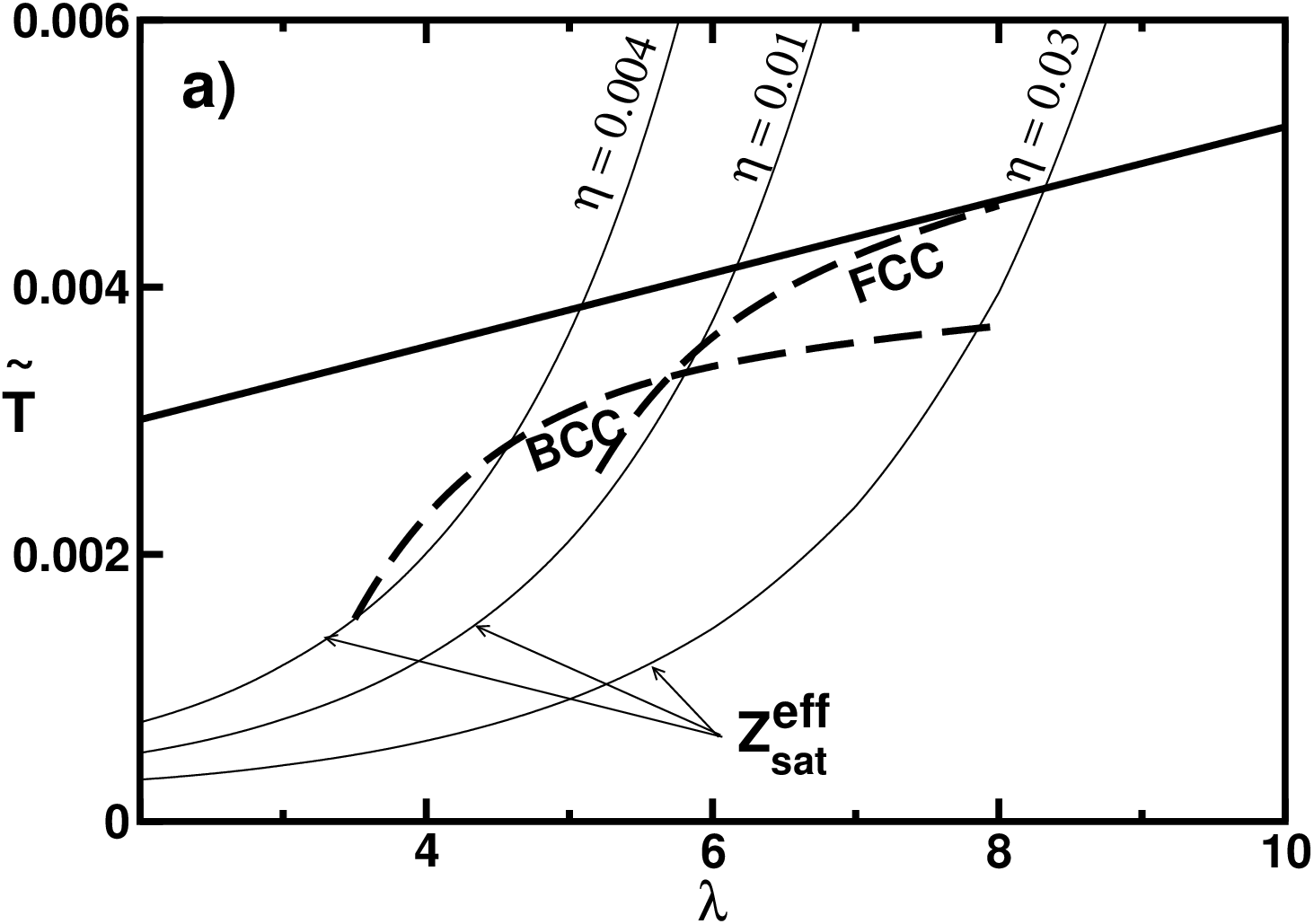}
\hfill \includegraphics[width=0.45\textwidth]{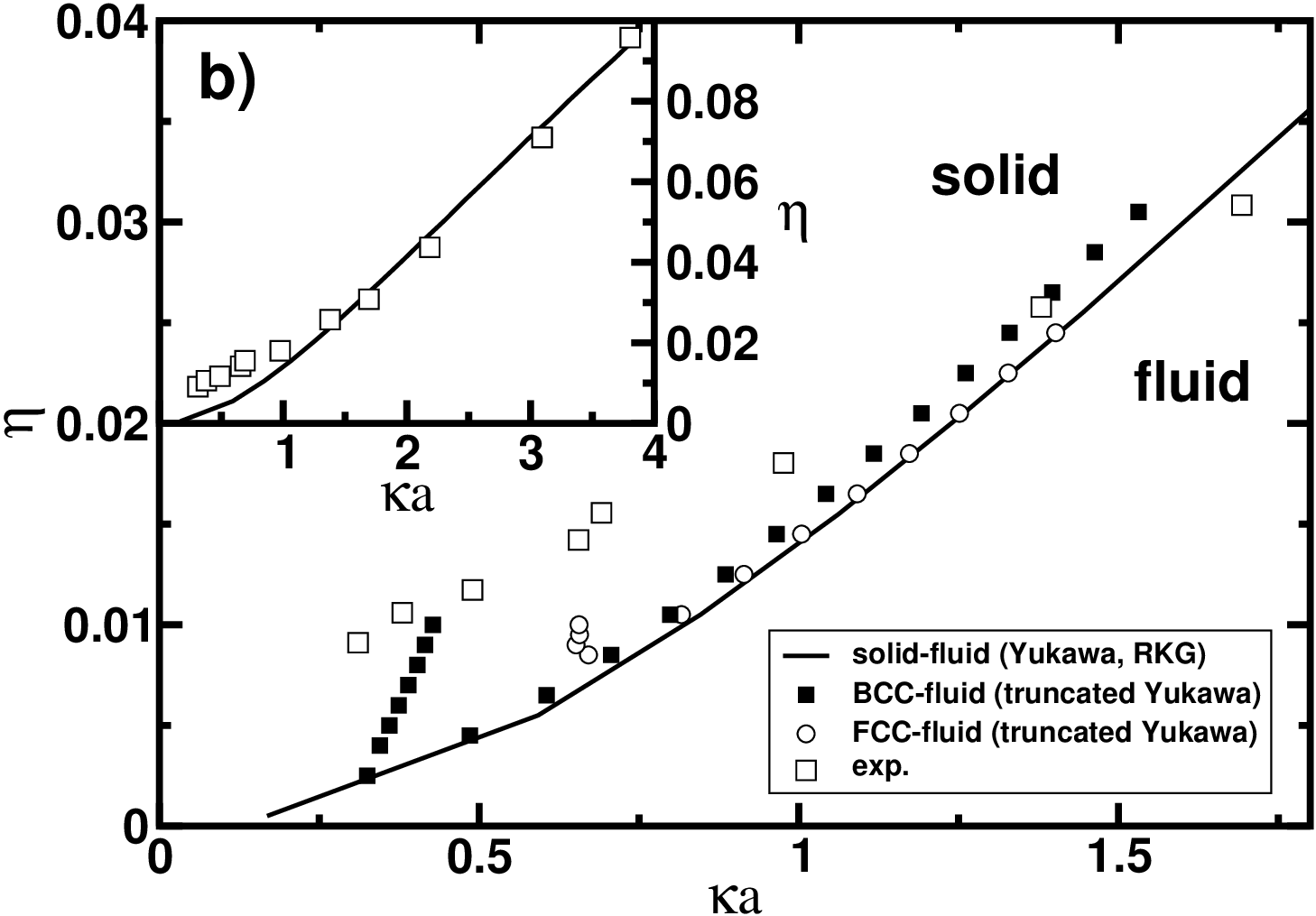}
\caption{\label{fig13} }
\end{figure}

\end{document}